\RequirePackage{fix-cm}
\documentclass[natbib,smallextended]{svjour3}       % onecolumn (second format)
\smartqed  % flush right qed marks, e.g. at end of proof
\usepackage{graphicx}
\usepackage{color}
\usepackage{epsfig,amsmath}
\usepackage{times}
\usepackage{aas_macros}
\newcommand{\gamray}{$\gamma-$ray}

 \journalname{svjour3}
\begin{document}

\title{Radio emission from Supernova Remnants
\thanks{G. Dubner and E. Giacani are members of the ``Carrera del Investigador Cient\'\i fico" of CONICET, Argentina}
}


\titlerunning{Radio Supernova Remnants}        % if too long for running head

\author{Gloria Dubner         \and
         Elsa Giacani %etc.
}

\authorrunning{Dubner \& Giacani} % if too long for running head

\institute{G. Dubner \at
              Instituto de Astronom\'\i a y F\'\i sica del Espacio (IAFE, CONICET-UBA), Casilla de Correo 67, Suc. 28, 1428 Buenos Aires, Argentina \\
              Tel.: +54-11-4789-0179\\
              Fax: +54-11-4786-8114\\
              \email{gdubner@iafe.uba.ar}           %  \\
           \and
           E. Giacani \at
           Instituto de Astronom\'\i a y F\'\i sica del Espacio (IAFE, CONICET-UBA), Casilla de Correo 67, Suc. 28, 1428 Buenos Aires, Argentina \\
                           \email{egiacani@iafe.uba.ar}     
}

\date{Received: date / Accepted: date}

\maketitle

\begin{abstract}

The explosion of a supernova releases almost instantaneously about 10$^{51}$ ergs of mechanic energy, changing irreversibly the physical and chemical properties of large regions in the galaxies. The stellar ejecta,  the nebula resulting from the powerful shock waves, and sometimes a compact stellar remnant, constitute a supernova remnant (SNR). They can radiate their energy across the whole electromagnetic spectrum, but the great majority are radio sources. Almost 70 years after the first detection of radio emission coming from a SNR, great progress has been achieved in the comprehension of their physical characteristics and evolution. We review the present knowledge of different aspects of radio remnants, focusing on sources of the Milky Way and the Magellanic Clouds, where the SNRs  can be spatially resolved. We present a brief overview of theoretical background, analyze morphology and polarization properties, and  review and critical discuss different  methods applied to determine the radio spectrum and distances. The consequences of the interaction between the SNR shocks and the surrounding medium are examined, including the question of whether SNRs can trigger the formation of new stars. Cases of multispectral comparison are presented. A section is devoted to reviewing recent results of radio SNRs in the Magellanic Clouds, with particular emphasis on the radio properties of SN 1987A, an ideal laboratory to investigate dynamical evolution of an SNR in near real time. The review concludes with a summary of issues on radio SNRs that deserve further study, and analyzing the prospects for future research with the latest generation radio telescopes.

\end{abstract}

%Please provide an abstract of 150 to 250 words. The abstract should not contain any undefined abbreviations or unspecified references.\\
%Please provide 4 to 6 keywords which can be used for indexing purposes.\\
%Please use the keywords defined by the editors of Astronomy \& Astrophysics, The Astrophysical Journal and Monthly Notices of the Royal Astronomical Society. The list of allowed keywords can be found on the web pages of these journals.
%\\
%Insert your abstract here. Include keywords, PACS and mathematical
%subject classification numbers as needed.

\section{Introduction}
\label{intro}
In 1919, the Swedish astronomer Knut Lundmark put forward the hypothesis that ``along with ordinary novae, on rare occasions stars flare up that are tens of thousands of times as bright at maximum". Zwicky and Baade proposed in 1934 that such stars be called {\it supernovae}, and though in the  opinion of Shklovsky  the name was rather absurd, it was rapidly popularized and now universally designates an event of stellar explosion \citep{shklovsky78}.

\begin{figure}[th!]
\centering
\includegraphics[ width=12cm]  {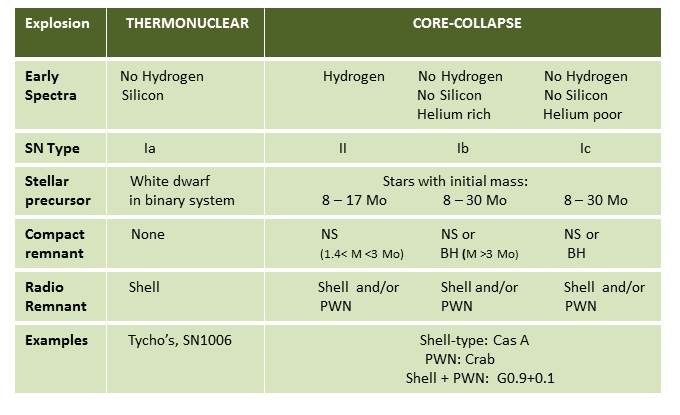}
\caption{Diagram summarizing the types of SN explosion, their precursors and possible radio remnants. NS is the acronym of neutron star, BH of black hole and PWN of pulsar wind nebula. Mass limits are taken from \citet{Smartt2009}. Mass quoted for NS and BH correspond to the compact core }
\label{diagram}
\end{figure}

Supernove (SNe) can be broadly classified into two big groups depending on the explosion mechanism: the Type Ia  are the result of a  runaway thermonuclear explosion of a degenerate carbon-oxygen stellar core (most likely a white dwarf in a binary system). The specific progenitor systems and the processes that lead to their ignition have not yet been clearly identified for these SNe as recently reviewed by \citet{Maoz2014}. The other big group of SNe involves  Types Ib, Ic and II, which are the product of gravitational core-collapse of massive stars  (initial mass M$\geq$ 8M$_{\odot}$) that have exhausted all their nuclear fuel. If the  stellar core contains between 1.4 and  3 M$_{\odot}$,  the  compact remnant is a neutron star, while if the mass of the collapsed core is larger than 3 M$_{\odot}$, a black hole is formed. As summarized by \citet{Smartt2009} there is a diversity of evolutionary scenarios, where metallicity, binarity, and rotation may play important roles in determining the end states, what kind of compact remnant leave, etc. 
As a general introduction to the radio SNRs and their connection with the explosion mechanisms that give birth to them, Fig. \ref{diagram} summarizes the types of SNe, their precursors and possible remnants. 

Independently of the type of explosion,   about 10$^{51}$ ergs of mechanic energy are suddenly deposited  in the
interstellar medium and several tens of solar masses of stellar material are ejected. The outer layers of the star blow off in all directions and the enormous explosion power imparts great velocities to the portions of the expelled envelope. 
   A nebula is formed that expands at a speed that can reach 5000 to 10000 km s$^{-1}$. This high expansion velocity is the main sign that distinguishes the remnants of supernova outbursts from other nebulosity.  Hundreds  or thousands of years later, the ejected material will begin to be slowed down by the ambient medium, the velocity will start to fall  to hundreds or even tens of km s$^{-1}$, and ultimately will disperse and merge with the surrounding gas. But through thousands of years a distinctive nebula, the supernova remnant (SNR) persists and can, eventually, be detected across the whole electromagnetic spectrum  from radio to gamma-rays. The neutron stars, the compact objects left by most of the supernova explosions, will continue radiating energy for many more millions of years.  

Before the advent of radio astronomy only two SNRs were known, the Crab Nebula and Kepler's SNR. Radio observations played a crucial role in the discovery and investigation of SNRs and their environs, revolutionizing the knowledge in the field.

In 1948, at the very opening era of radio astronomy, the British astronomers Sir Martin Ryle and Graham Smith detected an unusually bright radio source, Cassiopeia A (Cas A). The flux of this source was comparable to the radio flux of the quiet Sun at meter wavelengths, and it seemed to come from a place where no optical emission was apparent. Faint, filamentary optical emission  was only  later discovered by Baade and Minkowski \citep{Baade1954} from plates acquired in 1951 with the 200-inch Palomar telescope.    But soon after 1948 several radio sources associated with SNRs were found in our Galaxy: the Crab Nebula, the remnants of the Tycho  and Kepler supernovae, and a filamentary nebulae in the constellation of Cygnus. Then arose the fundamental question of the nature of those radio waves, since their radiation had nothing in common with thermal black body radiation, but rather was more akin to the nonthermal  Galactic radiation detected in the early years of the radio astronomy  by Karl Jansky and Grote Reber \citep{Jansky1933, Reber1944}. There was clearly a complete disagreement between the observed radio spectra of SNRs and those of thermal radio sources.  The correct idea
explaining radio emission of SNRs was proposed in 1950 by Alfv\'en and Herlofson \citep{alfven50}, and independently by Kiepenheuer
\citep{kiepenheuer50}. The answer was synchrotron radiation. The synchrotron nature was later confirmed by the detection of polarization in the Crab Nebula by \citet{mayer57}.

In the following years, radio continuum surveys carried out at different frequencies were the main tool for identifying  new SNRs through their non-thermal spectrum, the ``fingerprint" in this energy band. Whenever the instruments improve in sensitivity and angular resolution, new sources are found and this field is in continuous progress. From the very first lists of SNR candidates  published by \citet{Aizu1967} and \citet{Poveda1968} (with 25 SNRs listed),  superseded by  \citet{Milne1970} (who increased the list to 94 members in the Milky Way and 3 in the Large Magellanic Cloud),  \citet{Downes1971}, \citet{Ilovaisky1972}, \citet{Clark1976},  numerous works reported the discovery of new sources and refined their classifications. A  step forward came from the contribution of  \citet{Brogan2006}, who based on high-quality low-frequency radio observations disclosed the presence of 31 new SNRs in the inner Galaxy, increasing in about 15\% the total number of Galactic SNRs known by then. 
At the present time there are 294 firmly classified SNRs in our Galaxy, as compiled and permanently updated by   \citet{Green2014}\footnote{http://www.mrao.cam.ac.uk/surveys/snrs/}, of which the large majority ($\sim 95$ \%) are radio sources. Also, \citet{Ferrand2012} keep an updated census of high energy observations of Galactic SNRs\footnote{http://www.physics.umanitoba.ca/snr/SNRcat/}, listing their physical properties along with a summary of observations of these remnants with various X-ray and \gamray\ observatories.

It has to be noted that  after years of searching, the total number of detected SNRs in the Milky Way is significantly smaller than expected. A number at least 3 times greater of SNRs is statistically predicted on the basis of OB stars count, pulsar birth rates, SN rates in other Local Group galaxies, and predicted lifetime of radio synchrotron emitting sources  in the sky.  Such deficit is generally attributed to selection effects \citep{Brogan2006}, when old, faint, large remnants, as well as young, small sources, remain below the  threshold in sensitivity and/or spatial resolution of the large Galactic surveys performed up to now.

\section{A brief overview of the theory needed to understand radio emission from SNRs}
\label{sec:theory}

Many good text books present the physical background of synchrotron emission in the radio astronomical context \citep[e.g.][]{Pacholczyk1970, moffet75,verschuur88,harwit88,rohlfs90, rohlfswilson96}, and more recently a good synthesis is presented  in the review of SNRs at high-energies by \citet{reynolds08}. Here we summarize just the basic results needed for the interpretation of observational data with a brief guide to infer intrinsic properties and physical conditions of a source from the observed parameters. Readers not interested in the background aspects can jump to Section \ref{sec:morphology}.

\subsection{Synchrotron radiation:}
 A single relativistic electron moving in an external magnetic field B, will emit continuum spectrum radiation. The frequency near which the emission reaches its maximum intensity is called the critical frequency for synchrotron emission and is  defined as: 
\begin{equation}
\nu_c = \frac {3e} {4 \pi m c} B_\perp \left( \frac {E} {m c^2}\right)^2 = C_1 B_\perp E^2
\end{equation}
where the constant C$_1 = 6.266 \times 10^{18}$ in cgs units or C$_1 =$ 16.08 when $\nu$ is in MHz, B in microgauss and E in GeV \citep{moffet75}. For example, if we observe radiation at $\sim 1500$ MHz and assume a typical B field of $\sim$ 10 $\mu$G, we can conclude that such radiation was produced by electrons with an energy of 3 GeV. It can be shown that the time required for an electron to lose half of its energy is $t_{1/2} = C_3 / (B^2~ E_0) $ where $C_3 = 8.35 \times 10^9$ yrs when B is in $\mu$G and E in GeV. Then the electron of our example would radiate half of its energy in about 30 million years. For the Crab Nebula, with B$\sim$ 500 $\mu$G the time  to lose a substantial portion of its energy for an electron radiating in radio (500 MHz) is $\sim$ 100000 yrs, for an optical photon (600 nm) about 100 yrs, and for X-rays (4 keV) $\sim$ 2.4 yrs. Such short lifetimes (much shorter than the age of the Crab Nebula) implies that the pulsar is permanently  feeding energy into relativistic electrons.

Actually relativistic particles are not alone but in an ensemble. The synchrotron radiation that an observer detects from a particular volume element  of a radio source comes from all the electrons with the same pitch angle. We can assume that the source consists of a volume V containing a tangled magnetic field with average strength B, in which there are electrons with an energy density with power law distribution (as the empirical evidence of the cosmic rays close to the Earth shows) $ n(E)dE = n_0E^{-\gamma}dE$  between a range of energies E$_1 \leq E \leq E_2$ to avoid divergences in the extremes. In this case it can be shown that the emitted radiation has a spectrum $\epsilon_\nu \propto \nu^{-\frac{(\gamma-1)}{2}} = \nu^{-\alpha}$, with ${\alpha}$ = $\frac {(\gamma -1)} {2}$  the emission spectral index  . It can be easily concluded that the power law of the SNRs radio spectrum simply reflects the power law energy spectrum of the relativistic particles responsible for the radio waves that we observe. After carrying out the corresponding calculations, it can be shown that the flux density of a synchrotron source with a volume V follows the expression:
\begin{equation}
 S(\nu)  = 0.017 a(\alpha) V B^{(\alpha+1)} \left( \frac{6.26 \times 10^{18}} {\nu {\mathrm (Hz)}}\right)^{\alpha}  \mathrm {Jy},
\end{equation} 
where $a(\alpha)$ takes values  like 0.283 for a spectral index $\alpha= 0$ ; 0.103 for $\alpha= 0.5$; or 0.085 for $\alpha= 0.75 $ \citep{rohlfs90}.
The convention adopted here is that the flux density S$_\nu \propto \nu^{-\alpha}$ following \citet{rohlfs90}.

 During the evolution of a remnant the magnetic field strength will decrease producing a secular decay of the flux density for young radio sources \citep{Shklovsky1960a}. The exact rate depends on the field configuration, but assuming that the magnetic flux remains constant, B(R) is proportional to R$^{-2}$ for an adiabatic expanding nebula of radius R. Then, the flux density should decrease with R as S$_{\nu} \propto$ R$^{-2\gamma}$, which can be expressed as a function of time as S$_{\nu} \propto$ t$^{-4\gamma/5}$. Therefore, it is expected that the flux density decreases with time as:

\begin{equation}
\dot{S}_{\nu} / S_{\nu} = -\frac{4\gamma}{5t}.
\end{equation}
This result is a rough approximation for a young source evolving into a uniform medium. A more rigurous result can be obtained by numerically computing the radio emission along with the hydrodynamic evolution of the SNR. In some cases for very young remnants (less than $\sim$ 100 yrs), the radio emission can follow the opposite behavior and increase the flux with time. This is the case for example of the youngest SNR  observed in our Galaxy, G1.9+0.3, whose  flux density increased by a factor of 1.25 over 13 yr at 1.4 GHz \citep {Green2008}, and the SN 1987A in the Magellanic Cloud where the flux density increases in an exponential way with an average annual rate (measured at the year 20 since the SN explosion) of the order of 15\% at different radio frequencies \citep{zanardo2010}.

\subsection{Particle acceleration:}

An important question is what accelerates the particles to relativistic speeds in a SNR. If the remnant contains a pulsar, the answer is clear, it is the pulsar which supplies freshly accelerated electrons over the full lifetime of the SNR. But what happens with SNRs that do not have a neutron star in their interior? If the amount of material picked up by the supernova 
shock is large compared to the ejecta mass, then the particles can be accelerated in the shock waves.
 
The basis of all theories to explain  acceleration mechanisms in shock waves is that particles can gain energy in collisions with irregularities of the magnetic field.  \citet{Bell1978a, Bell1978b} and \citet{Blandford1978} proposed that the most efficient process is diffusive shock acceleration (DSA), where electrons gain energy after multiple crossing through a shock wave \citep{Fermi1949}.  For energetically unimportant test particles which do not influence the flow structure (the so-called test-particle limit),  the synchrotron spectral index is fixed by the shock compression ratio r by $\alpha = 3/ (2(r-1))$. In the case of strong shocks with a compression ratio of 4, this mechanism predicts a spectral index $\alpha$ = 0.5 for the radio flux density S.  Coincidences and departures from this theoretical prediction are discussed in Section \ref{sec:spectra}. For extensive discussions and reviews of standard and non-linear particle acceleration  theories,  see for example \citet{Blandford1987, reynolds08, Malkov2001, Reynolds2011, Jones2011, Urosevic2014}, and references therein.

\subsection{Magnetic field:}

It was early theoretically  suggested by  \citet{vanderlaan1962}  and later reinforced by \citet{Whiteoak1968} through a polarization study,  that  the magnetic field responsible for the synchrotron emission observed in SNRs  was the  interstellar field  compressed by the explosion. The hypothesis was plausible because even a relatively weak compression of the few micro Gauss of the ambient magnetic field, would result in an observable radio source due to the strong  dependence  of  volume  emissivity  on  the  magnetic  field  strength.  

Years later,  however,  indirect observational arguments, like the  thickness of the X-ray rims in the SNRs  Cas A, Kepler, Tycho, RCW 86 and SN 1006 \citep[e.g.][and references therein]{Jun1996,  Parizot2006, Ballet2006}, and the rapid variability (on timescale of few years) observed in X-ray spots in G347.3-0.5 (SNR RXJ1713.7-3946) \citep{Uchiyama2007} and in Cas A \citep{Patnaude2007}, pointed to the fact that the  magnetic fields must be  from tens to hundreds of times more intense than expected from adiabatic compression of the interstellar magnetic field, and  some extra mechanism of amplification is required. Such mechanism must have little connection with the past of the SNR, since  the seven studied cases have different progenitors, different explosion types (SNe types II and Ia), and  contain or not a central neutron star, etc., but all of them require considerable magnetic field amplification to explain the observations. 

Several mechanisms have been proposed that amplify the magnetic field. \citet{Schure2012} and \citet{Reynolds2012} present reviews on observations and various theories on  magnetic field amplification  in the presence of a cosmic ray population, showing that  cosmic ray streaming can induce instabilities that act to amplify the magnetic field.  Some of the proposed models include amplification due to flow instablities (of the Rayleigh-Taylor or Rankine-Hugoniot class) between the mean flow and clouds in the circumstellar and/or interstellar medium \citep[e.g.][]{Jun1995, Jun1996}; amplification  as a result of very efficient acceleration of nuclear cosmic rays at the outer shock \citep[][ and references therein]{Volk2005}; turbulent amplification driven by cosmic-ray pressure \citep{Beresnyak2009, Drury2012};  nonlinear DSA \citep[][ and references therein]{ Schure2012}, etc.  However, the global problem about the origin, properties and evolution of the magnetic field in SNRs, is still a matter of debate.

\subsection{The energy stored in particles and in magnetic fields:}
For an SNR, it is of interest to determine the energy content. This can be done by integrating over the electron spectrum between the energies $E_1$ and $E_2$. If evidence of a cutoff is observed at one end or  the other of the radio spectrum, the total electron energy can be expressed in terms of the cutoff frequencies $\nu_1$ and $\nu_2$ (the critical frequencies for the cutoff energies) and $\alpha$, the spectral index of the source. By assuming that each electron radiates only at its critical frequency, we obtain an expression for the electron energy. 
\begin{equation}
 U_e = \frac{L~C_1^{1/2}} {C_3 B^{3/2}}~ \frac{ (2 - 2\alpha)}{(1 - 2\alpha)} ~\frac{\nu_2^{1/2 - \alpha} -\nu_1^{1/2 - \alpha}} {\nu_2^{1 - \alpha} -\nu_1^{1 - \alpha}} ~~~  \rm{( if} ~\alpha \neq  {0.5}~~ \rm{or}~ 1) 
\end{equation}
where L is the total luminosity of the source derived from the observed flux density S ($ L = 4 \pi d^2 S$ assuming isotropic radiation and that the distance $d$ to the source is known), the constant C$_1$ is as defined above, C$_3= 2.368 \times 10^{-3}$ in cgs units and B is the average strength of the magnetic field. The magnetic field may not be measured directly, but its value may sometimes be inferred (at least an order of magnitude as is shown below). Observations at different frequencies provide an estimate of the spectral index $\alpha$.

The lower cutoff energy can be taken equal to the electron rest-mass energy (an electron that is not relativistic will not produce synchrotron radiation). Observations show departures from a power-law spectrum in the range from tens to few hundreds of MHz. Therefore a lower cutoff frequency of 10$^7$ Hz is often assumed. For the upper cutoff, if the source under study has X-ray measurements, the break frequency derived from the intersection of radio and X-ray spectral slopes can be used as upper frequency for radio emission. If there is no other indication, it is customary to use $\sim 10^{10} ~\rm{or} ~10^{11}$ Hz as the upper cutoff.

The energy of a source that emits synchrotron radiation can be mainly contained in two forms, as kinetic energy of the relativistic particles $U_{part}$ and as energy stored in the magnetic field $U_{mag}$ (thermal radio radiation is negligible, therefore thermal energy will be small compared to the formers, and  gravitational energy is probably the source of particles and magnetic energies, but almost nothing is known about it and is not considered). 

To estimate the energy stored in particles, we have to take into account that in addition to electrons there must be protons and other energetic baryons in the radiating source (though heavy particles emit negligible amounts of radiation because they are accelerated much less by the Lorentz force). Therefore, $U_{part} = \eta U_e$, where $\eta$ is a factor that takes all other particles into account. From various methods, $\eta$ has been estimated to be 50 (the value from cosmic rays near Earth)  or 100 \citep[from models applied to radio sources in our Galaxy, e.g.][]{burbidge59}. Fortunately, as shown below, this uncertainty does not have a strong effect on the  fundamental conclusions.

On the other hand, the energy stored in the magnetic field for a source of volume V is: $U_{mag} = VB^2/8\pi$. Therefore, 
\begin{equation}
U_{tot} = U_{part} + U_{mag} = \eta A L B^{-3/2} + V B^2/8 \pi,
\end{equation}
 where A takes into account the constants $C_1, C_3$ and the shape factor in frequency. 

It can be noticed that the energy stored in particles, proportional to $B^{-3/2}$,  will dominate for small field strengths,  while the magnetic energy ($ \propto B^2$) dominates for large fields. In consequence, the total energy $U_{tot}$ has a minimum for a magnetic field intensity
\begin{equation}
B(U_{min}) = \left(\frac{6 \pi a A L}{V}\right)^{2/7},
\end{equation}
for which $U_{part} / U_{mag} = 4/3 $.  Thus, the minimum total energy to produce the observed radio emission corresponds quite closely to an equipartition between relativistic particles and magnetic fields.  
If the magnetic field is exactly that of equipartition, then $ U_{min} = 0.5 (\eta A L)^{4/7} V^{3/7}$. In this way, we can obtain a firm lower limit of the energy requirements of a synchrotron source. Examples of the use of these relations can be found in \citet{Frail1996} for the PWN in W44, \citet{dubner2000} for the SNR W28, or in \citet{castelletti2007} for W44. 

The relations  presented above are useful to get first estimates of energy content and magnetic field strength in radio SNRs, but it should be made  clear that there is no physical justification
for the energy components of the source being close to equipartition. It has been conjectured that motions in the plasma may stretch and tangle the magnetic fields  and  turbulent motions
might also accelerate particles to high energies, thus taking  the plasma closer to equipartition, but these are only conjectures and, in fact, these radio sources might be far from equipartition.

\subsection{Dynamical evolution:}
Since the separation of the SNR evolution in four distinct  phases  originally proposed by \citet{Woltjer1972}, the scheme was maintained up to the present and is generally defined as follows:  
\begin{itemize}

\item { \it  Free expansion phase} This phase occurs when the shock wave created by the explosion moves outwards into the interstellar gas at highly supersonic speed and compressed interstellar gas accumulates behind
the strong shock front.  This  material is separated from the ejected stellar material by the contact discontinuity (a surface between two different materials with similar pressure and velocity but different density). Behind the contact discontinuity, a reverse shock starts to form in
the ejected stellar material. After some time, the accumulated mass of the ISM compressed between the forward shock and the contact discontinuity equals the ejected mass of stellar material, and it starts affecting the expansion of the SNR, marking the beginning of:

\item {\it The adiabatic expansion or Sedov-Taylor phase} Here, the expansion is driven by the
thermal pressure of the hot gas. In this stage, after the passage of the reverse shock, the interior of the SNR is so hot that the energy losses by radiation are very small. Since all atoms are ionized, there is no recombination, and  the cooling of the
gas is only due to the expansion. This stage has an exact self-similar solution \citep{Taylor1950, Sedov1959}. Later,  as the SNR expands and cools adiabatically, it reaches a critical temperature of about $10^6$ K, the ionized atoms start  capturing free electrons and they can lose their excitation energy by radiation.  The radiative losses of energy become significant,
setting the end of the adiabatic expansion of the SNR. The efficient radiative cooling decreases the thermal pressure in the post-shock region  and the expansion slows down.
The SNR is entering in the so-called:

\item {\it Snow plough or radiative phase} when more and more interstellar gas
is accumulated until the swept-up mass is much larger than the ejected stellar material. Finally, the shell breaks up into individual pieces, probably due to a Rayleigh-Taylor
instability (hot thin gas pushes cool dense gas) and the SNR goes into the final phase.

\item {\it Dispersion}, as the expansion velocity decreases to values typical of the interstellar gas and the SNR  disperses into the ISM. 

\end{itemize}

This scheme is an oversimplification and, as it was pointed out by \citet{Jones1998} distinct phases may be brief, may not occur at all, or may occur simultaneously in different regions of the same remnant. A detailed description of the hydrodynamical evolution of SNRs can be found in \citet{Lequeux2005}.

\citet{Chevalier1974} developed a set of numerical models to describe the
spherically symmetric expansion of a SNR in a uniform medium, providing
semi-analytical expressions to estimate expanding radius, age, shock
velocity, etc. for different initial and ambient conditions. Although this
work was later complemented with many different variations to consider
inhomogeneous surroundings, interaction with circumstellar matter, etc.
\citep[e.g.][]{Chevalier1982b, Truelove1999}, it is still a classic that
can be used as a first approach to derive physical parameters from observed
properties in SNRs.  For example, for an initial energy $E_0=3 \times 10^{50}$ ergs,
ambient density $n_0=1$ cm$^{-3}$ and $B_0=3 \times 10^{-6}$ Gauss, the radius and shock velocity of an SNR  at late times of evolution (after about $5 \times 10^4$ yrs) can be estimated within the model approximations using the following relations:  $R_0=21.9~ t_5^{-0.31}$ pc and 
 $v_{sh}=66.5~ t_5^{-0.69}$ km s$^{-1} $. 
\citet{Vink2012} presents a good overview of different
analytical models developed for SNRs expanding in various environments.

\section {Radio morphology}
\label{sec:morphology}

As already mentioned, the vast majority of SNRs in our Galaxy were first recognized from radio observations. In fact, out of 294 known remnants in our Galaxy, only 20  have not been either detected in the radio band, or are poorly defined by current radio observations \citep{Green2014}. The morphology and brightness distribution in SNRs contain important information about the nature of the SNR and its possible hydrodynamical evolution. 

\begin{figure}[ht!]
\centering
\includegraphics[width=1\textwidth] {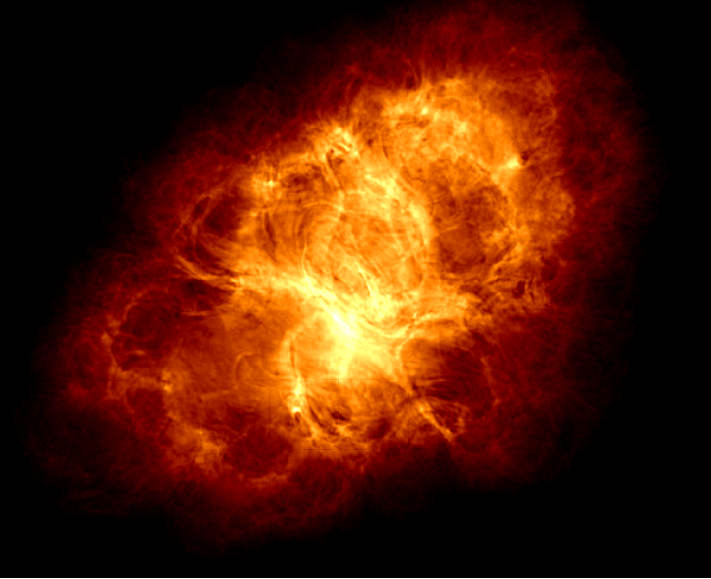}
\caption{Detailed radio image of the Crab Nebula, obtained at 3 GHz with the Karl Jansky Very Large Array (NRAO) in the A-configuration. The image has a HPBW of $0^{\prime\prime}.9 \times 0^{\prime\prime}.8$ and the rms noise is better than 30 $\mu$Jy (Dubner et al. 2015 in preparation)}
\label{crab}
\end{figure}

Supernova remnants have been traditionally classified based on their radio morphology into three broad categories \citep[see] [for an historical perspective on SNR naming conventions]{Weiler1988}. They are:
\paragraph {\it a) shell-type SNRs,} 
whose appearance is characterized by a limb brightened shell or ring formed initially by the ejecta from the SN, and  at later times also by swept up  surrounding material. The diameter of the shell corresponds to the expanding shock wave produced by the explosion. In this case the particles responsible for the observed synchrotron radiation are believed to be accelerated at the shock front.
\paragraph {\it b) filled-center or plerions,} in which the radio brightness is centrally concentrated and is often linearly polarized. In this case, the accelerated particles and magnetic fields responsible for the synchrotron
emission are injected by a pulsar created in the supernova event. This pulsar transfers the bulk of its rotational energy in a wind of relativistic particles and Poynting flux that interacts with the surrounding medium creating a synchrotron emitting nebula named ``Pulsar Wind Nebula" (PWN). 
PWNe can emit non-thermal radiation over the whole electromagnetic spectrum.
The Crab Nebula is a prime example of this class of remnants.  Figure \ref{crab} shows a high-quality, sub-arcsec resolution radio image of the Crab Nebula at 3 GHz, obtained with the Karl Jansky Very Large Array (JVLA, NRAO) as part of a project of simultaneous multi-wavelengths investigation of this source \citep[][Dubner et al. 2015, in preparation]{Krassilchtchikov2014}. See also \citet{Bietenholz2015} for an image of
Crab at 5 GHz with high resolution. It has to be noted that there are several  plerionic SNRs, for which no pulsar has yet been found. For these cases  the detection of the nebula is a strong indication that the powering
source must be an undetected pulsar. \citet{GaenslerSlane06} presented a complete review of PWNe as seen across the whole electromagnetic spectrum.
\paragraph{\it c) composite,} for which the SNR appears to have both a shell and  an internal non-thermal pulsar driven nebula.
As noted by \citet{Green2014} the term ``composite" has been also used by some authors to describe SNRs with radio shell and centrally brightened thermal X-ray emission. Such SNRs are also known as ``mixed-morphology" (M-M) SNRs \citep{rho1998}.  Since the physical nature of the  X-ray central component of these remnants is completely different from the non-thermal PWNe, we prefer to include them in a fourth morphological class. 
\paragraph{\it d) mixed-morphology (M-M), }  SNRs with synchrotron radio shell and  central  thermal X-ray emission. They  have attracted more attention in the last years since several of them appear related to molecular clouds and to \gamray\ sources, mostly of hadronic nature
(when relativistic particles collide with dense ambient gas producing neutral pions,  which decay into \gamray\ ). The catalog by \citet{Ferrand2012} lists  references for such associations. \citet{Vink2012} summarized their properties. Since new members have been reported in the last couple of  years, to keep an updated census of these remnants we list in Table \ref{MM} the  M-M SNRs in the Milky Way, together with their associations with molecular emission and \gamray s detections.

\begin{table*}
\renewcommand{\arraystretch}{1.0}
\centering
\caption{Mixed-Morphology SNRs}
\label{MM}
\vspace{0.26cm}
\begin{footnotesize}
\begin{tabular}{llccccl}
\hline\hline
SNR  & Name  &  Molecular  & OH (1720 MHz) & GeV & TeV & Ref\\
&  & Material& Masers &  & & \\ 
\hline
G0.0+0.0 & Sgr A East &  & Y &  &  & 1\\
G6.4-0.1 & W28 & Y & Y & Y & Y & 2,3,4, 5\\
G31.9+0.0 & 3C391 & Y & Y & Y &  & 3, 6, 7\\
G33.6+0.1 & Kes 79 & Y & & Y & & 8, 9\\
G34.7-0.4 & W44 & Y & Y & Y &  & 10, 11, 12\\
G41.1-0.3 & 3C397 & Y & & & & 13 \\
G43.3-0.2 & W49B & Y & & Y & Y & 14, 15, 16 \\
G49.2-0.7 & W51C &Y & Y & Y & Y & 17, 18, 19, 20 \\
G53.6-2.2 & 3C400.2 &  & & & &  \\
G65.3+5.7 & & & & & &  \\
G82.2+5.0 & W63 & & & & &  \\
G89.0+4.7 & HB 21 & Y & & Y& & 21, 22 \\
G93.7-0.2 & CTB 104A & & & & &  \\
G116.9+0.2 & CTB 1 & & & & &  \\
G132.7+1.3 & HB 3 & Y & & & & 23 \\
G156.2+5.7 & & & & & &  \\
G160.9+2.6 & HB 9 & & & & &  \\
G166.0+4.3 & VRO 42.05.01 & Y & & Y & & 24, 25 \\
G189.1+3.0 & IC443 & Y & Y & Y &Y  & 11, 26, 27, 28 \\
G272.2-3.2 & & & & & &  \\
G290.1-0.8 & MSH 11-61A  & & & & &  \\
G304.6+0.1 & Kes 17& Y & & Y & & 29, 30, 31\\
G327.4+0.4 & Kes27 & & & & &  \\
G337.8-0.1 & Kes 41 & Y & Y & & & 32, 33 \\
G344.7-0.1 & & Y & & Y & Y & 34, 35, 36 \\
G352.7-0.1 & & & & & & 37\\
G357.1-0.1 & Tornado & Y & Y & Y& & 38, 39 \\
G359.1-0.5 & & Y & Y  & & & 40, 41  \\
\hline
\end{tabular}
\end{footnotesize}
{References: (1) \citet{Yusef96}; (2) \citet{Reach2005}; (3) \citet{Frail96}; (4) \citet{Aharonian08}; (5) \citet{Hanabata14}; (6) \citet{ReachRho99}; (7) \citet{Castro10}; (8) \citet{giacani09}; (9) \citet{Auchettl14}; (10) \citet{Seta98}; (11) \citet{Claussen97}; (12) \citet{Abdo10a}; (13) \citet{Jiang10}; (14) \citet{Zhu14}; (15) \citet{Brun11}; (16) \citet{Abdo10b}; (17) \citet{Koo97}; (18) \citet{Green97}; (19) \citet{abdo09}; (20) \citet{Feinstein09}; (21) \citet{Koo01}; (22) \citet{Reichardt12}; (23) \citet{Routledge91}; (24) \citet{Huang86}; (25) \citet{Araya13}; (26) \citet{Su14}; (27) \citet{Ackermann13}; (28)\citet{Acciari09}; (29) \citet{Combi2010}; (30) \citet{Gelfand13}; (31) \citet{Wu11}; (32) \citet{Combi2008}; (33) \citet{Koralesky98}; (34) \citet{giacani2011}; (35) \citet {Abdo13}; (36) \citet{Aharonian06}; (37) \citet{giacani09}; (38) \citet{Lazendic04}; (39) \citet{Castro13}; (40) \citet{Hewitt08}; (41) \citet{Uchida92}.}

\end{table*}

\vspace{0.5cm}

In the recent compilation of Galactic SNRs by \citet{Green2014}, 79\% of remnants are classified as shell-type  (including the mixed-morphology), 12\% as composite, and 5\% as plerions. The rest of the remnants do not fit into any of the aforementioned conventional types, defying the simple spherically symmetric expansion solution. 

\begin{figure}[ht!]
\vspace{1cm}
\includegraphics[width=0.50\textwidth]{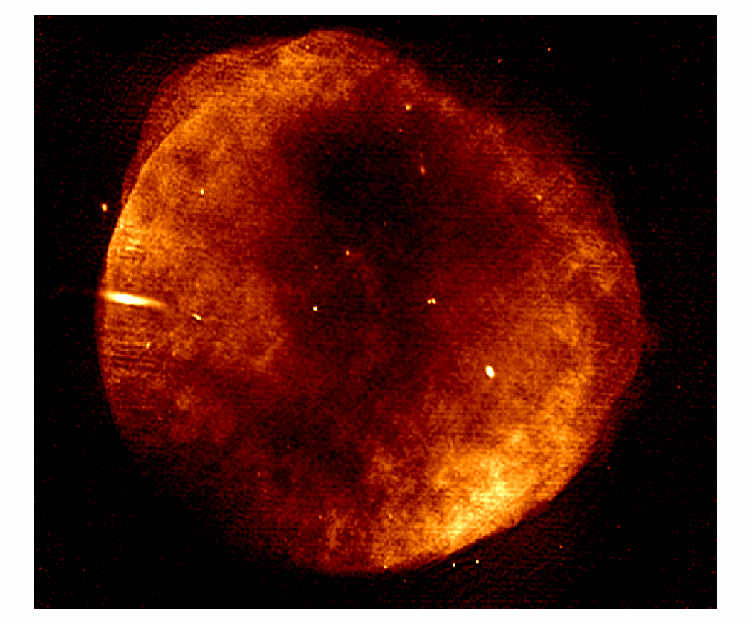}~\hfill
\includegraphics[width=0.45\textwidth]{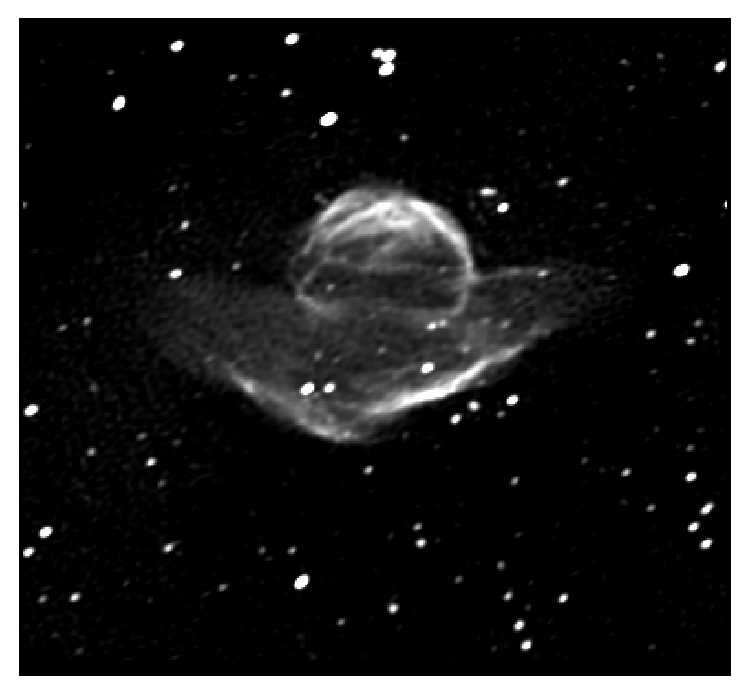}
\hspace{1cm}
\includegraphics[width=0.45\textwidth]{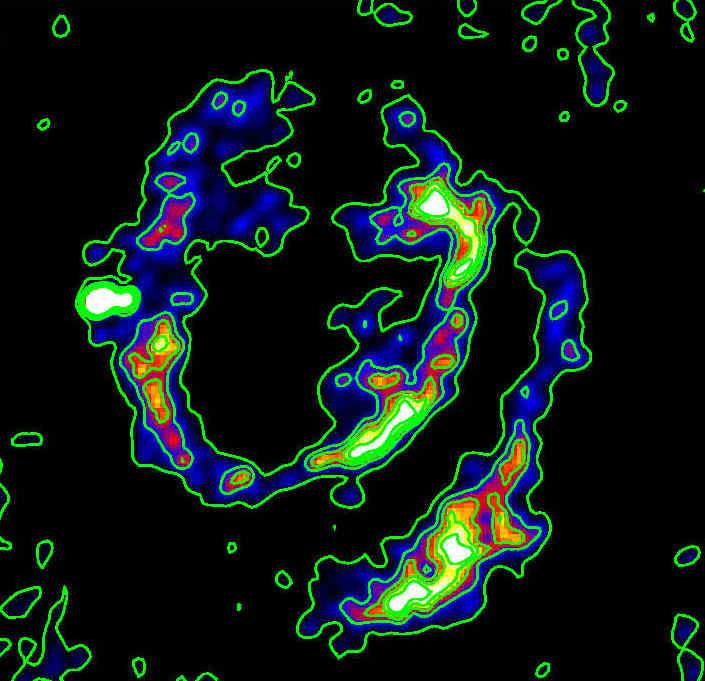}~\hfill
\includegraphics[width=0.55\textwidth]{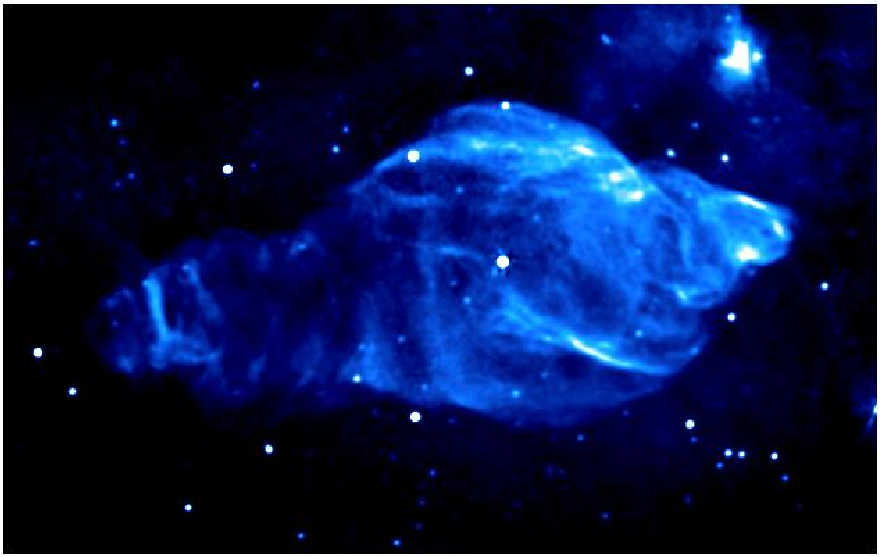}

 \caption{Radio images of SNRs with a variety of morphologies. {\it Upper left:} SN1006 at 1.4 GHz as taken from \citet{Petruk2009}; {\it Upper right:} VRO 42.05.01 at 1.4 GHz as extracted from the CGPS public database; {\it Bottom left:} G352.7-0.1 at 5 GHz from \citet{giacani09}; {\it Bottom right:} W50 at 1.4 GHz taken from \citet {Dubner1998}}
\label{morfologia}
\end{figure}

With regard to shell-type SNRs, multi-frequency observations of SNRs carried out in the last years with increasing resolution and sensitivity have demonstrated that the actual morphologies of these objects 
are highly complex and less than 20\% of the shell-types have an almost  complete circular ring appearance.  In effect, in radio waves SNRs exhibit an ample variety of shapes, such as  {\it blow-out:} in which, part of the shell appears to have expanded more rapidly than the rest.  An example of this type is  the SNR  VRO 42.05.01 (G166.0+4.3) (Fig. \ref{morfologia} {\it upper-right}) that at first sight  appears to be two different sources adjoining in the plane of the sky \citep{Landecker1982}, but that HI observations demonstrated that it is a single SNR breaking out from a warm medium into a tenuous interstellar cavity \citep{Pineault1987, Landecker1989}; {\it barrel-shaped or bilateral:} SNRs characterized by a clear axis of symmetry, low level of emission along this axis, and two bright limbs on either sides \citep{Gaensler1998}. Typical examples of this type are G296.5+10.0 \citep{giacani2000} and SN1006 \citep{Reynolds1986, Petruk2009} (Fig. \ref{morfologia} {\it upper-left}); {\it multi-shells:} remnants with two or more overlapping rings of emission. Examples of this type are G357.7$-$0.1 \citep{manchester87}, 3C400.2 \citep{dubner94}, G352.1$-$0.1 \citep{giacani09} (Fig. \ref{morfologia} {\it bottom-left}). 

Another interesting morphology is created when a central compact object injects into a SNR a  flow of particles along collimated {\it bi-polar jets}. This appears to  be the origin of the distorted elongated shape observed in some SNRs, as in the SNR W50 \citep[][Fig. \ref{morfologia} {\it bottom-right}]{Dubner1998}, in Puppis A \citep[Fig. 9 in][]{Castelletti2006}, and in Cas A \citep{Hwang2004}. In the case of the W50/SS433 system, the radio observations confirmed  the connection between the sub-arcsec relativistic jets from SS433 and the extended helical nebula over 5 orders of magnitude in scale \citep{Dubner1998}. A similar origin was proposed for the ``ears" observed in the SNR Puppis A, where the central neutron star would be responsible for the production of collimated outflows that impact on the shell. In the case of Cas A, the marked bi-polar asymmetry, in this case revealed in X-rays, was explained by  \citet{Hwang2004} as the result of opposite symmetrical  jets  produced deep within the progenitor of the SN explosion. The nature of this morphology is the subject of observational and theoretical efforts to explain it.

In the investigation of SNR morphologies, an additional complication is that the observed shape is a two-dimensional projection of a three-dimensional object and depends on their orientation with respect to the line of sight and projection effects. As a curiosity useful to illustrate this point, we reference  the animation of a 3-D vision of the optical emission of the Veil Nebula SNR  presented by J. P. Metsavainio\footnote{http://astroanarchy.blogspot.com}. An example of the complications inherent to projection effects when studying the origin of the radio morphology is the case of the Galactic SNR G352.7-0.1 (Fig. \ref{morfologia} {\it bottom left}) whose radio emission projected in the sky plane looks like  two concentric rings. This shape has been alternately proposed to originate in a ``barrel-shaped"  type of SNR \citep{giacani09} or from a blow-out scenario, where the SN explosion took place near the border of a molecular cloud \citep{toledoroy2014}.

\subsection {Can the SN type be constrained on the basis of the radio morphology of the SNR?}

This has been a challenging question for decades. The great diversity of shapes observed in radio SNRs, reflects not only different properties of the progenitor star and of explosion mechanisms, but also echoes the properties of the ambient magnetic field and the matter distribution in the circumstellar and  interstellar medium.  Disentangling the different causes is a complex task that remains as a central question. 

Several attempts have been made to infer the type of supernovae on the basis of the observed remnants. In the X-ray domain, \citet{Lopez2009, Lopez2011}  developed an observational method to characterize the type of explosion of young SNRs  by measuring global and local morphological properties  of the  X-ray line and thermal emission in numerous young SNRs in our Galaxy and in the Large Magellanic Cloud, finding that  the remnants of Type Ia SNe have statistically more spherical and mirror-symmetric thermal X-ray emission than SNRs coming from core-collapse origin. These studies were later extended to infrared (IR) morphology  by \citet{Peters2013} through the investigation of the symmetry of the warm dust emission, concluding again that Type Ia SNRs are statistically more circular and mirror symmetric than core-collapse SNRs. This is explained as due to different circumstellar environments and explosion geometries of the progenitors. Additional ways of identifying the type of SN have been proposed using the X-ray emission, distinguishing Ia remnants from core-collapse ones by virtue of their ejecta composition \citep{Hughes1995} (Fe-rich and O-poor SNRs are likely Ia, while SNRs dominated by O and Ne lines with weak Fe L emission are likely core-collapse ones). More recently, \citet{Yamaguchi2014} presented an observational diagnostic to discriminate between progenitor types based on the Fe K-shell X-ray emission. The authors find that in remnants from SN Ia, the Fe-rich ejecta is significantly less ionized than in remnants from core-collapse SNe.  Their results also indicate that there is a strong connection between the explosion type and the ambient medium density.

The morphological criteria that seems to work for X-rays and IR to infer the SN type from the SNRs characteristics are clearly useless for radio SNRs.  If we blindly apply the ``circularity"  criterion to Cas A and Tycho's SNR,  two circularly symmetric young radio remnants, both of them would be classified as SN Ia, but  Cas A is core-collapse and Tycho's is  SN Ia. On the other hand, the SNRs SN 1006 and G296.5+10.0  are two perfect examples of ``mirror-symmetric" sources, but SN1006  is a SN Ia, while G296.5+10.0 has a central compact object suggesting a core-collapse origin \citep{Gaensler1998, HarveySmith2010}.

One important reason that complicates the connection of a radio SNR with its precursor  is that while X-rays may retain information about the characteristics of the exploded star,  the complexity of the interaction between the shock front and the ejecta, circumstellar and interstellar matter, can soon mask this information in the radio emission.  Once the shock front sweeps up a certain amount of ambient  gas, the radio synchrotron emission ignores the explosion properties and it is mostly conditioned by inhomogeneities in the surrounding medium, hydrodynamic instabilities in the flow, turbulence behind the shock, effects of magnetic fields, etc. \citep[e.g.][]{Chevalier1982a, Chevalier1982b, Dwarkadas2005}. Sometimes, the traces left by mass loss episodes of the stellar progenitor can help to identify the class of supernova. The presence of a neutron star inside the remnant and/or the existence of the pulsar wind nebulae is an unquestionable evidence of a core-collapse event. The absence of it, however, does not prove anything because pulsars have high kick velocity \citep[e.g.][]{Hansen1997} and can be outside the SNR  far from the explosion site, also central compact objects can be radio silent, etc. An indirect indicator of core-collapse SN is  finding the SNR  very close to or inmersed in a molecular cloud that might be the birthplace of a massive star that ended its life as SN Ib,c or II.  

In summary, the radio morphology alone does not provide a useful tool to distinguish between different types of SNe.

\section{Polarization}

\label{sec:Polarization}

Radio polarization observations of SNRs provide essential information on the degree of order and orientation of the ordered component of the magnetic field, which themselves influence  the morphology of the remnants and the intensity of radiation. 

As the radio emission in SNRs is primarily synchrotron, the radiation is linearly polarized; therefore from the observed polarization electric vector, the direction of the orthogonally aligned magnetic field can in principle be determined.  However, in practice, the observed polarization can be highly  compromised by Faraday rotation since the electric field vector of the radiation is rotated during the propagation inside the SNR and the interstellar medium. In the simplest case, the angle of rotation $\psi$ varies proportionally to the square of the wavelength $\lambda$, such that $\psi$ = RM $\lambda^{2}$, with the constant of proportionality, or the rotation measure RM, defined as

\begin{equation}
{\rm RM (rad~m^{-2})} = 0.81 \int N(cm^{-3})B_{\parallel}(\mu G) \rm{dl (pc)} 
\end{equation}

where N is the thermal electron density and $B_{\parallel}$ the magnetic field component along the line of sight l, and the integral extends along the entire line of sight. The sign of RM is determined by whether $B_{\parallel}$ points toward or away from the observer. In general, observations at three or more wavelengths are necessary to measure the RM without ambiguity and determine the true position angles of the magnetic field lines. 

From the synchrotron theory, the intrinsic degree of linear polarization of the radiation emitted  from electrons in a uniform magnetic field is independent of the frequency and given by P= $(2\alpha + 2)/(2\alpha + 10/3)$, where $\alpha$ is the radio spectral index of the radiation \citep{Ginzburg1965, Pacholczyk1970}. For a non-thermal radio source with a spectral index $\alpha = 0.5$, the fractional polarization can  reach a maximum theoretical value of about $70 \%$, but in practice a much lower polarization percentage is usually observed.  

The reduction of the observed degree of polarization arises from different physical and instrumental effects, namely beam depolarization as a result of variations of the rotation measure  on spatial scales smaller than the antenna beam;  bandwidth depolarization, when the polarization angle changes across the receiver bandpass and the resulting non-parallel vectors are averaged; and  depths depolarization (also known as differential Faraday rotation) when there is superposition of emission from different depths along the line of sight, either internal or external to the radiating source,  which suffers  different Faraday rotation. Under these conditions, the dependence of $\psi$ with $\lambda^{2}$ is no longer valid. To mitigate  this effect several models have been proposed by,  e.g. \citet{Burn1966} and \citet{Sazonov73}. For a discussion of basic depolarization mechanism and instrumental effects, see also \citet{Gardner&whiteoak}, \citet{MilneDickel75}; \citet {Milne1980}, and \citet{Reich2006}.

Recently, a new tool, called RM Synthesis or spectropolarimetry, has been implemented to recover the polarization structure at multiple Faraday depths along a particular line of sight with the additional advantage that minimizes  n$\pi$ ambiguities and  bandwidth depolarization \citep [for a more detailed explanation see][]{brentjens05, heald09}. Its application has become viable thanks to  technical and computational advances. Most application of this technique have been for interstellar or extragalactic studies \citep [a good review can be found  in ][]{Beck2012} and its utilization for studying magnetic fields in SNRs is a very new field of research. \citet {HarveySmith2010} applied this tool to the SNR G296.5+10.0 using data taken with the Australia Telescope Compact Array at frequencies near 1.4 GHz. The Faraday rotation maps \citep[Figure 1 in ][] {HarveySmith2010} shows a highly ordered rotation measure structure with an anti-symmetric RM morphology across the remnant.  The authors proposed that the observed RM pattern is the imprint of an azimuthal magnetic field in the stellar wind of the progenitor star. The expansion of the remnant into such a wind can account for the bilateral morphology of G296.5+10.0 as observed in the radio and X-ray bands. 

Even if the radio observations of SNRs are carried out in optimum conditions to minimize the effects mentioned above, the observed degree of polarization in SNRs is still considerably lower than the maximum possible theoretical value. This is an indication that the magnetic fields are primarily disordered. In general, the polarization degree has been found to be between 10 and 15 $\%$ \citep [see references in ][]{ReynoldsGilmore93}, with higher values between 35 and 60 $\%$ in some few exceptional cases, as for example in some regions in the Vela SNR \citep{Milne1980}, DA 530 \citep{Landecker1999}, G107.5-1.5 \citep{kothes03}, and SN1006 \citep{reynoso13}.

\begin{figure}
\vspace{1cm}
\includegraphics[width=0.42\textwidth]{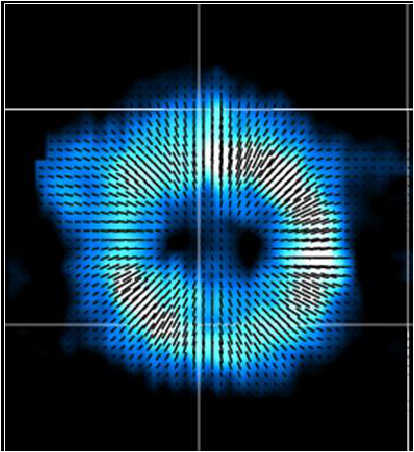}~\hfill
\includegraphics[width=0.60\textwidth]{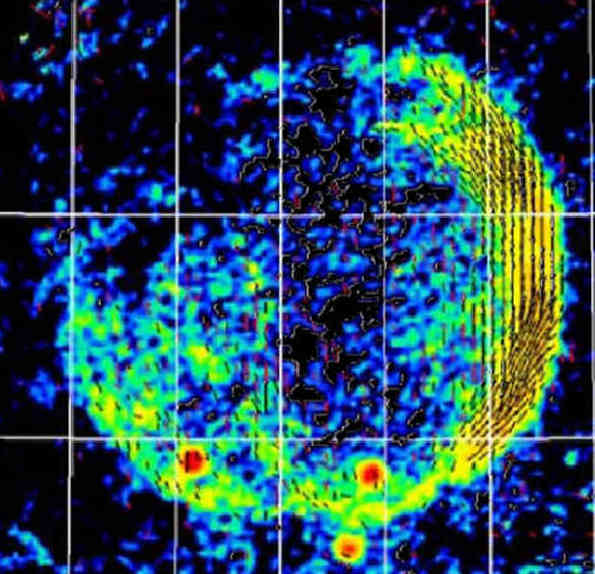}
      \caption{ Radio polarization bars in B-field direction as obtained with the Effelsberg 100-m telescope. {\it Left:} SNR Cas A at 32 GHz; {\it right:} SNR CTB1 at 10.55 GHz. (courtesy of W. Reich)}
\label{polarization}
\end{figure}

At large spatial scale, it has been proposed that the intrinsic orientation of the magnetic field in SNRs, as inferred from the radio observations, shows a typical pattern depending on their age. The earliest observations of the young SNR Cas A showed a convincingly radial magnetic field with respect to the shock front  \citep{Mayer68}, while in the case of the old Vela SNR, the radio polarization map showed a near tangential direction in the brighter emission \citep{Milne68}. Later on, polarization measurements  carried out by \citet{Milne87} over 27 SNRs confirmed that in young remnants the alignment of the magnetic field is predominantly in the radial direction, whereas the dominant orientation of the field in older remnants is parallel to the shock front or tangled. Subsequent observations have supported this picture \citep {Landecker1999, furst04, Wood08}. In Figure \ref{polarization}, we show the intrinsic magnetic field distribution in the SNR Cas A (left) and in CTB1 right), illustrating two extreme cases of  magnetic field distribution predominantly radial and  tangential, respectively.

\begin{figure}
\begin{center}
\includegraphics[width=0.7\textwidth]{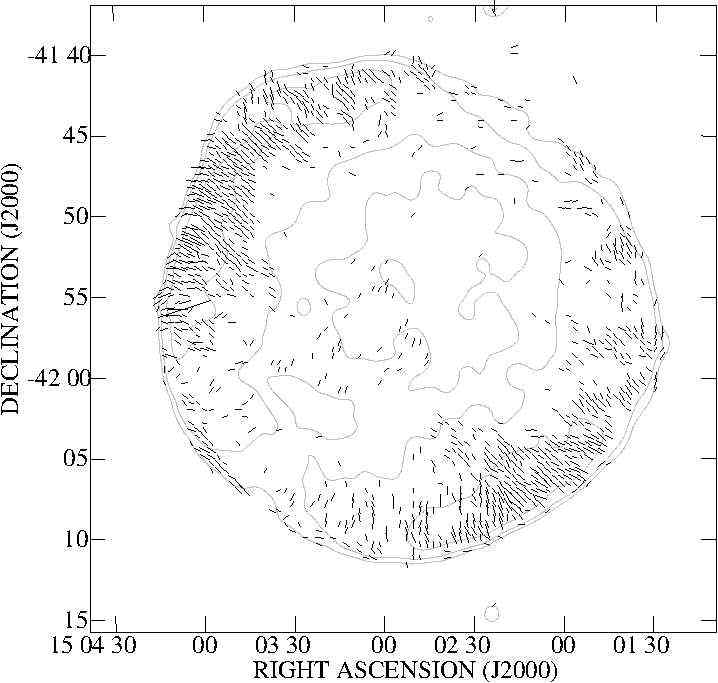}
      \caption{Magnetic field distribution on SN 1006 at 1.4 GHz \citep[from][]{reynoso13}}
\label{SN1006}
\end{center}
\end{figure}

The general consensus on the tangentially ordered fields observed in older remnants is that they originate by compression in radiative shocks with large shock compression ratios. Regarding the radial component, however,  the origin is still controversial. It has been generally  attributed to stretching of Rayleigh-Taylor (R-T) instability fingers that occurs at the contact discontinuity which separates the  ejecta from the shocked, more tenuous, circumstellar medium (CSM) \citep [e.g.,][]{Jun1996}. However,  polarization observations of SN1006 \citep{reynoso13} show radial orientation near the forward shock (Fig. \ref{SN1006}) . This radio polarization study was carried out using data obtained with the VLA and ATCA instruments at 1.4 GHz with an angular resolution of about 10$^{\prime\prime}$. 
The high quality of these data highlights the complex structure of the magnetic field distribution.  A notable characteristic in SN1006 is that even when the orientation of the magnetic field vectors across the SNR shell appears to be radial, a large fraction of the magnetic
vectors lie parallel to the Galactic plane (whose orientation is perpendicular to the symmetry axis of the bright lobes). \citet{reynoso13} conclude from this evidence that the ambient magnetic field must be roughly parallel to the Galactic plane, and the SNR retains some knowledge of the orientation even after the passage of the shock front.   Besides, while the degree of polarization in the two bright radio lobes of SN1006 is about 17\%,  a value  as high as about 60\% is found  towards the faint SE border of the remnant. In brief, the brightest radio, X-ray and TeV features,  the NE and SW lobes of SN1006, have the lowest polarization fractions (indicating the presence of a disordered, turbulent magnetic field), while  in the SE where the synchrotron emission is faint, the polarization is high (ordered field). In this case, the authors conclude that  the most efficient particle acceleration occurs for shocks  in which the magnetic field direction and shock normal are quasi-parallel, while inefficient acceleration and little to no generation of magnetic turbulence are obtained for the quasi-perpendicular case, a result with implications for DSA theories.

Over the past several years, a number of survey projects were launched in order to map the polarized emission from the Galaxy in great detail. Single-dish as well as synthesis telescopes located in the northern and southern hemispheres are being used. An updated list of the surveys of the Galactic polarized emission that have been conducted and being carried out, can be found in Tables 1 and 2, respectively, in \citet{Landecker12}, while a review of the radio polarization measurements from the beginnings to the present day was presented by \citet{wielebinski12}. As an important by-product, it can be mentioned that these surveys together with the total intensity are the main source for identifying new SNRs. Recently, 79 SNRs have been observed during the Sino-German polarization survey of the Galactic plane performed with the Urumqi 25-m telescope at 6 cm \citep{han2014}. Combining these data with observations made with the Effelsberg 100-m telescope at 21 and 11 cm, for the first time traced the magnetic field orientation for 23 out of 79 sources observed and identified two new SNRs, G178.2-4.2 and G25.1-2.3.

\section{Radio spectra}
\label{sec:spectra}

\begin{figure}
\vspace{1cm}
\begin{center}
\includegraphics[width=0.8\textwidth, clip=true, bb=16 30 624 450]{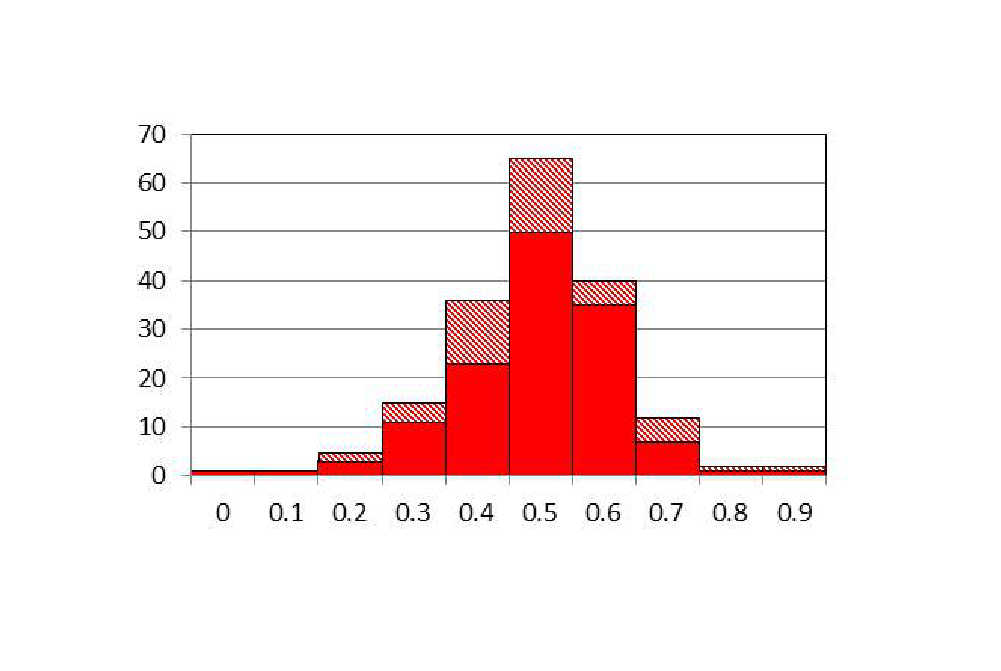}
~\hfill
      \caption{Histogram of spectral indices of  shell-type Galactic SNRs as extracted from \citet{Green2014}. The solid bars correspond to firm spectral index determination, while the shadowed ones to SNRs with uncertain spectral indices}
\label{spectralindices}
\end{center}
\end{figure}

The knowledge of the spectral behavior of SNRs is vital to understanding  particle acceleration at the shock fronts and the role of SNRs as factories of Galactic cosmic rays. The study of the global and  spatially resolved radio-continuum spectra of SNRs  provides significant constraints on shock acceleration theories.  As mentioned in Section \ref{sec:theory}, in the case of strong shocks with a compression ratio of 4,  DSA (diffusive shock acceleration) predicts for the radio flux density a  spectral index $\alpha$ = 0.5.   However, when compared with the observed radio spectral indices, it is found that only 50 out of the 294 SNRs listed in \citet{Green2014}'s catalogue have  $\alpha$ between 0.5 and 0.6, a number that only increases to 65 if the doubtful or poorly determined spectral indices are also considered.  Figure \ref{spectralindices}  is an update of the summary of spectral indices presented by \citet{Reynolds2011}, after selecting the SNRs classified as  shell-type  and separating those sources with well known spectral index from those listed with a question mark, for which a simple power law is not adequate to describe their radio spectra.   The members of the composite and mixed-morphology   classes are included as shell-type radio SNRs.  It is worth mentioning that among the SNRs classified as shell-type, almost 44\%  have poorly determined  spectra, with a percentage as high as 54\% for sources located in the fourth Galactic quadrant, very likely due to observational selection effects. These numbers, however, have to be considered with caution as the tabulated spectral indices in Green's catalogue  come from very diverse studies, with a wide range of (uncatalogued) errors.

Nevertheless, the fact that less than one fourth of the catalogued Galactic SNRs have the theoretically expected spectrum poses a problem to theorists. Different explanations have been proposed to address this. For instance, flatter spectra for increasing energy  can be produced by shocks with low Mach number. The drawback of this hypothesis is that only a  few SNRs would be expected to have such slow shocks, not enough to explain all the observed cases \citep{Reynolds2011}. Also, second-order Fermi (stochastic) acceleration can play a role (Ostrowski 1999) and non-linear  shock acceleration (when accelerated particles influence the shock dynamics) can also produce flatter spectra \citep{Ellison1991, Reynolds1992}.  

Another important aspect to note is that theory predicts that the particle acceleration must be very efficient in young SNRs, corresponding to free-expansion and especially early Sedov phases of evolution. This effect should translate in flatter  spectra  ($\alpha \leq 0.5$) for young SNRs. However,  observations show that young SNRs have steeper indices (e.g.  $\alpha= 0.77$ for Cas A, 0.58 for Tycho's SNR, 0.6 for SN1006,  and about 0.8, with extreme values in the range 1.1 to 0.3 for the remnant of SN1987A.

In addition, the spectra of SNRs frequently show departures from a power law  at the lower radio frequency extreme due to  extrinsic or intrinsic processes that modify the electron energy distribution or the radio propagation. Three different physical processes  that can affect the low radio frequency extreme have been proposed. They are: thermal absorption, synchrotron self-absorption, and the ``Tsytovitch effect''. Thermal absorption occurs when the non-thermal SNR radiation traverses a region with thermal plasma, producing a  low-frequency turnover in the spectrum. Synchrotron self-absorption takes place if the intensity of synchrotron radiation within the source becomes sufficiently high,  then reabsorption of the radiation through the synchrotron mechanism may become important, modifying the spectrum at low frequencies. 
The Tsytovitch  effect can only be important in sources with very small size and weak magnetic fields \citep [see][for details] {moffet75}. \citet{Bell2011}  proposed that the spectrum can either steepen or flatten depending on the angle between the shock normal and the large-scale upstream magnetic field. 

To narrow the gap between theory and observations and understand the physical meaning of the spectra,  it is important to get good spectral index maps of many SNRs expanding in different environments and, if possible, in different evolutionary stages, to compare young and old population. The  accurate radio spectral study of SNRs provides three different pieces of important information: (1) the {\it global index}, a key parameter to constraining particle acceleration theories as mentioned above, (2) the {\it curvature} of the spectrum, useful to test  radiation mechanisms and also to separate intrinsic from extrinsic factors that may modify the spectrum, and (3) the existence of {\it local variations} within the remnant, that help to localize the sites where particle acceleration is taking place, the possible existence of radio PWNe (especially useful for the search of radio counterparts of PWNe discovered in other spectral regimes, like X- or $\gamma$- ray), and the presence of superimposed or embedded sources of thermal absorption.

Since intrinsic or extrinsic spectral variations are often subtle, an important key is obtaining a large enough leverage arm in frequency space to tease them out. In the spectral studies it is essential to separate the contribution of three factors that usually overlap hiding the investigated properties: 
 the intrinsic characteristics of the explosions, the  contribution of the environment, and observational selection effects often imposed by the type of instrument used to acquire the radio data. This last one is not a trivial issue, since interferometric observations can over resolve the source, missing total flux information, but single dish observations that produce accurate total flux measurements often confuse SNRs with nearby objects. An additional related issue is the problem of the proper subtraction of background emission, a serious issue since  low-intensity emission extending over large angular distances (usually close to the Galactic plane)  can result in a substantial contribution to the flux density of large SNRs. When analyzing the available radio spectra of SNRs, it is evident that most remnants are represented by only a few data points with substantial error bars, especially at the lower frequencies. In what follows, we discuss the different techniques developed to overcome the various problems that can affect the spectral studies and some interesting results.

\paragraph{Global spectrum}
To accurately determine global indices and their curvature, one  technique is  the T-T plot 
\citep{Leahy1998}, that fits pixel by pixel a linear relation to the brightness temperatures measured at two frequencies,  deriving a temperature index from the slope  ($T_\nu = T_0 \nu^{-\beta}, \rm{where} ~ \beta=\alpha + 2$). To investigate the possible curvatures, three or more frequencies must be used taking pairs of maps. This procedure  guarantees that the derived spectral index, either total or local,  is not sensitive to differences in the zero levels between the maps at the two considered frequencies. In any case, it is important that the images to be compared have similar angular resolution and sensitivity, whatever is the technique used to obtain the data (single-dish, interferometers or both combined).

\begin{figure}
\vspace{1cm}
\includegraphics[width=0.5\textwidth]{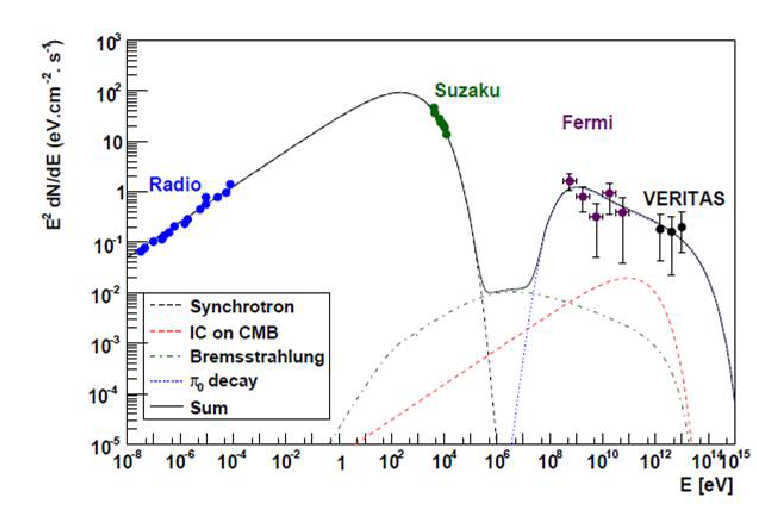}\hfill
\includegraphics[width=0.5\textwidth, clip]{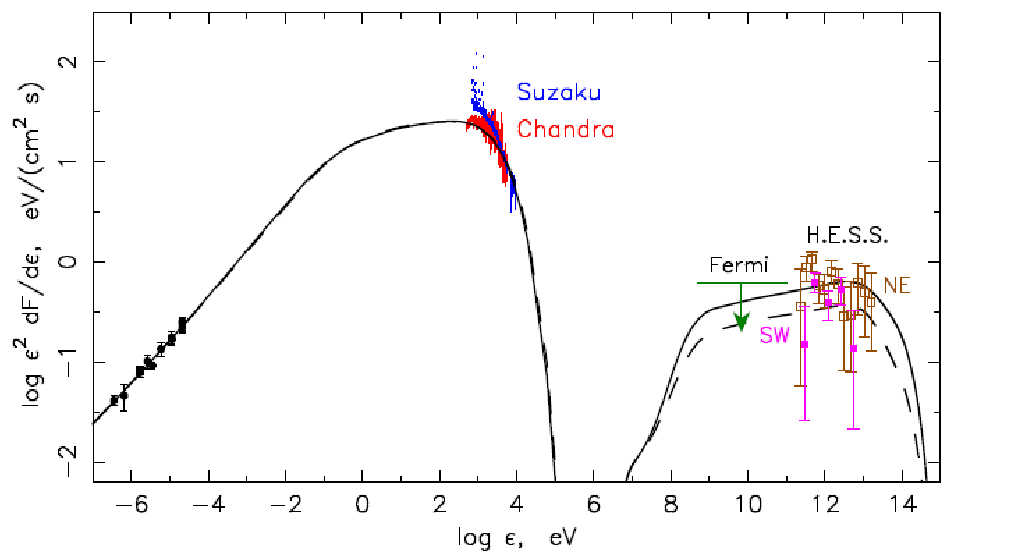}\hfill
\centering
\includegraphics[width=0.4\textwidth]{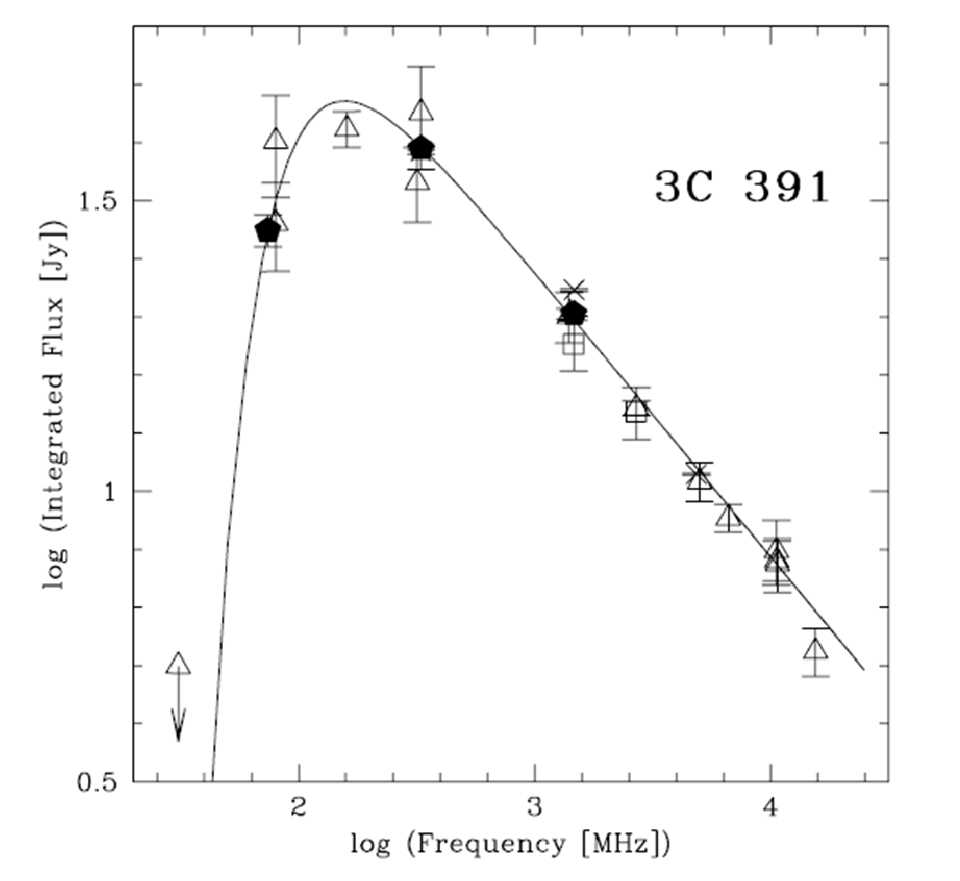}~\hfill

      \caption{{\it Upper} broadband spectral energy distribution  from radio to \gamray\  showing the importance of the radio data to fit models in Tycho SNR 
 \citep [{\it left}, from][]{Giordano2012} and in SN1006  \citep [{\it right}, from][]{Berezhko2012}. {\it Bottom} radio continuum spectrum for SNR 3C 391 showing a turnover in the spectrum at frequencies $\leq$ 100 MHz, indicative of free-free absorption from thermal ionized gas along the line of sight \citep [from][]{Brogan2005}}
\label{spectral-fit}
\end{figure}

\citet{Kassim1989} carried out a pioneer work of observing numerous SNRs of the first quadrant at low radio frequencies and compiled existing observations, producing an atlas of radio SNR spectra, finding that about 10 out of the 32 observed SNRs show turnovers at low radio frequencies (below $\sim$ 100 MHz), which was interpreted as due to the presence of a widespread, but inhomogeneous ionized absorbing medium along the line of sight, probably associated with the extended HII region envelopes (EHEs). Figure \ref{spectral-fit} shows an example of spectrum with low frequency turnover (bottom), and two examples of broadband spectral energy distribution (SED) to illustrate the utility of the radio spectra to fit models to the \gamray\ emission (upper).

\paragraph{Spatial spectral variations}

\begin{figure}
\centering
\includegraphics[width=0.5\textwidth]{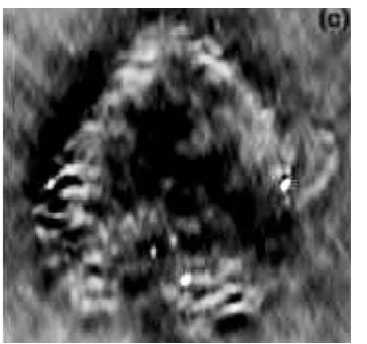}
      \caption{Tomographic image of Puppis A from \citet{Castelletti2006} for $\alpha_{\rm{test}} = 0.6$. Bright regions have $\alpha$ steeper than $\alpha_{\rm{test}}$, while dark regions correspond to $\alpha$ flatter than $\alpha_{\rm{test}}$}
\label{tomography-puppis}
\end{figure}

The study in radio wavelengths of spatial spectral variations inside a SNR can provide important information about the physical conditions of the interior plasma. One problem in the study of spectral variations across  an SNR is the confusion with unrelated overlapping structures. To address this difficulty, \citet{Katzstone1997} developed the tomographic technique, designed  to isolate structures with different spectra. Spectral tomography involves making a gallery of maps using different test $\alpha$, in which each tomographic image between frequencies $\nu_1$ and $\nu_2$  is $ I_t(\alpha_{\rm{test}}) = I_1 - (\nu_1 / \nu_2) ^{\alpha_{\rm{test}}} I_2$. Features or regions which have a spectral index identical to the test value will vanish in the tomographic map. Spatial components that have different spectral indices will appear as positive or negative features depending upon whether the spectrum is steeper or
flatter than the assumed test value. This method has been successfully applied to analyze spatial spectral variations, for example  in the SNRs of Tycho \citep{Katzstone2000}, Kepler \citep{DeLaney2002}, G292.0+1.8 \citep{gaensler03}, and in Puppis A \citep{Castelletti2006}.  The method is particularly useful to  locate small-scale spectral variations, providing  a more accurate picture than the simple comparison between images at different frequencies.  Figure \ref{tomography-puppis}  shows the tomographic image obtained for Puppis A  for $\alpha_{\rm{test}} = 0.6$, where it can be noticed a spectral pattern formed by short horizontal, almost parallel fringes 
with $\alpha$ alternatively steeper and flatter than the background, reproducing the  ``wave-like" morphology observed  in the total power image along the NE, NW, and S borders of Puppis A. 

A significant breakthrough occurred with the first spatially resolved detection of thermal absorption towards the SNR 49B \citep{Lacey2001}, helping to explain the long mysterious presence of radio recombination lines  towards this non-thermal source. 

Soon after, a number of additional cases of resolved thermal absorption came to light, as for example  the observation of a cocoon of thermal material in the SNR 3C391, marking the ionized boundary of the interaction of a SNR with a neighboring molecular cloud \citep{Brogan2005}. 

Another interesting example is the detailed spatial spectral study performed in Cas A, where  \citet{Kassim1995} showed  that the data were consistent with absorption by ionized gas inside the radio shell, probably related to unshocked ejecta still freely expanding within the boundaries of the reverse shock as delineated by X-ray observations.  \citet{Delaney2014}  extended these observations and their analysis to constrain the physical properties of a component of SNRs that has been heretofore very difficult to study. 

The spectral study carried out by \citet{Castelletti2011a} in IC443 revealed the existence of two different spectral components, both with flat spectrum ($\alpha \leq 0.25$) but of distinct origin, one extrinsic and the other intrinsic. One of these components coincide with the brightest parts of the remnant along the eastern border and perfectly matches the region where ionic lines are detected in the J and H  infrared bands. Such correspondence is the manifestation of the passage of a  J-type shock across an interacting molecular cloud that dissociated the molecules and later ionized the gas. This ionized gas produces thermal absorption along the line of sight, resulting in the observed flattened spectrum. The other flat spectrum component is more fragmented and located near the center of the SNR, in spatial coincidence with a region with a molecular cloud and \gamray\ emission  (Fig. \ref{IC443-comparison}), suggesting that the origin in this case is the particle acceleration that takes place at the shock front.

\section {Distances to SNRs}
\label{sec:distances}

A precise estimate of the distance to SNRs is essential to determine the physical parameters and understand their nature. It is, unfortunately, one of the most  difficult quantities to measure with accuracy. \citet{Wilson1970} expressed that ``At present, distances  to  only  a few SNRs are known with any reliability. Optical estimates are restricted to those SNRs which are rather near the Sun and, while nearly one hundred SNRs have been found (Milne 1970), 
distance estimates independent of any theory of SNR evolution are available for only a few sources". Today, more than four decades later the situation have scarcely improved and  only one-third of the catalogued Galactic SNRs has a distance estimate, several of them with large uncertainties or barely expressed as limits.

The most precise determinations come from the combination of measurements of different independent quantities. \citet{Trimble1973} addressed the distance calculation for the Crab Nebula using 12 different lines of evidence through methods depending on the dynamics of the nebular expansion, on physical processes in the supernova or the remnant, and  involving the properties of the interstellar medium between us and the nebula or the pulsar. This was an exceptionally well-studied case, and yet today it is affirmed that the true distance of Crab Nebula remains an open question owing to uncertainties in every method used to estimate its distance and the error  amounts 25\% (d=2 $\pm$ 0.5 kpc).

In general, for less famous sources, several methods have been used to measure distances to Galactic SNRs, including the  $\Sigma-$D (radio surface brightness-to-diameter) relation, kinematic distances through atomic and molecular absorption and emission, association with other objects with known distances (usually HII regions, OB associations, pulsars) assuming that they are neighbours, X-ray and optical observations, etc. Each approach has its advantages and disadvantages, briefly discussed in what follows.

\paragraph{$\Sigma - $D relation}

Although the application of this method is very controversial, it is described for its historical value and because in some cases it may provide a first guess when no other distance indicator is available. It is only applicable to shell-type SNRs. The basic assumption is that since at a first approximation $\Sigma_\nu$, the mean radio surface brightness at frequency $\nu$, can be assumed as an intrinsic property of the SNR, it is a distance-independent observational parameter \citep{Shklovsky1960a, Shklovsky1960b}.  The first theoretical $\Sigma - $D relation was derived by \citet{Shklovsky1960a} and the first empirical relations were derived by \citet{Poveda1968}. \citet {Clark1976}  describe that after reaching a maximum value shortly after the birth of the supernova, it may be expected that $\Sigma$ decreases monotonically  with time, while the outer linear diameter, D, of the expanding SNR will increase monotonically with time. The basic idea is then to construct an empirical $\Sigma - $D relation using calibrators with distances known by other methods. It is expected that  there is not too much scatter in this ``evolutionary track" .
For the relation  $\Sigma_\nu = A D^\beta$,  the distance is derived as d $ \propto \Sigma_\nu^{1/\beta}\theta^{-1}$ or, in terms of observable quantities, $ d \propto S_\nu^{1/\beta} \theta^{-(1+2/\beta)}$, where $S_\nu$ is the flux density at the observing frequency $\nu$ and $\theta$ is the angular diameter of the remnant. This formula also assumes that the SNR boundary is circular; when this is not the case it is commonly defined an equivalent diameter as $\theta = 2\sqrt{area/\pi}$. As mentioned, departures from sphericity are not uncommon as they depend on environmental conditions and/or the orientation of the observer  relative to the source. The ellipticity  of typical remnants  suggests that the variation of mean apparent angular diameter with orientation is sometimes as great as 10\%.

The reliability of this method has been seriously objected, since  there is no theoretical basis for a single equation to cover the evolution of SNRs over their whole lifetime and expanding in a variety of ambient media \citep[see discussions in][]{Green1984, Green1991}. Evolutionary paths may differ substantially from one SNR to another since they come from different stellar progenitors,  probably experienced explosions with distinct physical characteristics, and may be expanding in interstellar media with a wide range of densities, from the evacuated interior of a  wind-blown bubble, to a dense molecular cloud. 
However, in spite of the serious objections, this method has been  widely applied for years, and there have been several attempts to  improve it  in different ways. The distances of the sources used as calibrators were refined by changing the kinematical estimates through the application of a  more modern rotation curve for the Milky Way, by taking new data from associations of SNRs with pulsars or molecular clouds, or by comparing with new X-ray and optical data. The different estimated power law index $\beta$ varies in the range -2.38 to -5.2 \citep[see][for a summary and comments on the  different calibrations]{Case1998}. Pavlovic et al. (2014) presented the most recent fitting based on a revised calibration sample consisting of 65 Galactic SNRs, deriving $\beta = -5.2$. \citet{Arbutina2004} showed that the only case were a single $\Sigma-$D correlation could be established was for the SNRs in the starburst galaxy M82, and suggested that a similar behavior could exist for a sample of Galactic SNRs associated with large molecular clouds. They conclude that the density of the  environment is probably a crucial parameter to regulate the diameter during the SNR evolution.

\paragraph{Kinematical method}

Basically, this method is based in the construction of an absorption HI  spectrum by subtracting an average spectrum obtained  from an area projected against a region with  strong continuum emission  of the target SNR, from a spectrum, or average spectrum, of an adjacent background region. Using the radial velocity of the absorption peak and by applying a Galactic circular rotation model\footnote{ Updated models for the Milky Way are presented  in Bhattacharjee et al. (2014)} to convert radial velocities into distances, it can be set a distance, or range of acceptable distances, for the SNR.  Complications arise when the HI emission is patchy and cause spurious absorption features or  when the continuum is faint and it is not possible to construct acceptable absorption spectra, with peaks noticeable at least at  a 5-$\sigma$ level.  \citet{Tian2008} apply a combination of HI absorption plus CO emission to overcome some of the basic problems. Kinematical distances can also be derived from the study of the ISM around the SNR. When there is evidence of interaction between the SNR and the surrounding atomic or molecular gas (Section \ref{sec:interaction}), the radial velocity at which the best signature is identified can be used to establish an approximate distance to the SNR.

However, even when the radial velocity of the absorption or emission associated feature is clearly defined, there still exists the problem of the ambiguity  of the Galactic rotation curve within the Solar circle, i.e. for each radial velocity there are two corresponding distances equally spaced on either side of the tangent point. Moreover, although with the help of other indicators a single distance can be established, the circular rotation model still adds large intrinsic uncertainties (the model holds only for  low  Galactic latitudes and ignores  systematic stream motions in the Galaxy, rolling motions in the Galactic arms, random cloud velocities,  etc.) and the distances derived using kinematical methods have intrinsic (inherent to the method)  uncertainties greater than $\sim$ 25\% - 30\%. 

\paragraph{Distances derived from X-ray observations}

\citet{Kassim1994} proposed a formula for shell-type SNRs  to estimate distances in the cases where  it can be assumed that  the SNR shell is in the adiabatic expansion phase, that the measured X-ray 
temperature gives a reliable estimate of the SN shock velocity, and that an initial energy of the order of  E$_0 \sim 10^{51}$ erg is valid  for all SNe. In these cases,  the distance to a SNR could be derived as a function of the initial energy,  the observed angular diameter of the SNR shell, the measured X-ray flux corrected for interstellar absorption, the thermal temperature of the X-ray emitting gas, plus a function that describes the power emitted by hot electrons in a low-density plasma via free-free emission (that depends of  both the energy band of the emission and the temperature of the plasma) as  $D_s = 8.7 \times 10^6 \epsilon_0^{0.4} P(\Delta E,T)^{0.2} \theta^{-0.6} F_{X_0}^{-0.2} T^{-0.4} $. 
The method was applied to a sample of SNRs detected with ROSAT,  with the major uncertainty related to the basic physical assumptions  rather than to measurement errors.

Other application of  X-ray data to constrain  distances to SNRs was for example the one used in the study of  the SNR RX J1713.7-3946 (G347.3-0.5) first discovered by its X-ray emission \citep{Cassam-Chenai2004}. The procedure consists of plotting the cumulative absorbing column  calculated from atomic and molecular gas observations as a function of the radial velocity and comparing with the absorbing column NH as derived from a fit to the X-ray data. The radial velocity at which these quantities are equal is later translated to distance through a kinematic model of Galactic rotation.

In summary, the distances estimated for SNRs have to be considered with caution. Even in the best cases when more than one independent distance determination can be used, the derived distances are still imprecise, either because of observational inaccuracies or because of doubtful assumptions involved in the formulae used to calculate them.

\section{Interaction of SNRs with the surrounding ISM}
\label{sec:interaction}

The interaction between the strong SN blast wave and the surrounding interstellar medium has profound consequences on the remnant as well as on the gaseous interstellar matter. SN explosions are the main way of chemical enrichment of the ISM, while the distribution and physical conditions of the surrounding gas represent the primary physical constraint to the expansion of the SN shock.  The investigation of the SNR/ISM interaction  is not only necessary to improve our knowledge of SNRs, but also to understand the response of the interstellar gas to the enormous injection of energy and momentum that a SN explosion represents. Additionally, as already mentioned in Sec. \ref{sec:distances}, the identification of  physically associated interstellar gas serves to calculate the kinematical distance to the remnant. The study is also very important to understand the production of \gamray s, and hence the role of SNRs powering Galactic cosmic rays, as clouds illuminated by the protons accelerated in a nearby SNR could be bright \gamray\ sources \citep[e.g.][] {Aharonian1996, Berezhko2000, Gabici2009, Butt2009,Ellison2011}. 

The ISM  is formed by  various  components in different  physical conditions. It is arranged in a variety of structures and phases, including big complexes of dense clouds, hot  bubbles, sheets, walls, filaments, etc. \citep[e.g.][]{Lequeux2005, Cox2005}.  Some of these structures are directly created by the SNe and SNRs. The environmental characteristics play a crucial role in the shape, energetics,  temporal evolution and destiny of SNRs. Before analyzing the consequences of the mutual  interaction between SNRs and the ISM, in what follows we briefly review the basic characteristics of the ISM in our Galaxy.

\paragraph {Phases of the interstellar medium:} 

\begin{itemize}

\item {\it The molecular medium (MM)}, characterized by cold dense molecular clouds which are mostly gravitationally bound, with typical  temperatures  $\leq$ 100 K, volume densities $\geq 10^3  $ cm$^{-3}$, and volume filling factor  f $\leq$ 1\%.  A mass of about $1.5 \times 10^9$ M$_\odot$ of molecular gas is distributed predominantly along the spiral arms of the Milky Way,  occupying only a very small fraction of the ISM volume and within a narrow midplane with a scale height Z $\sim$ 50 to 75 pc.

\item {\it   The cold neutral medium (CNM)}, distributed in rather dense filaments or sheets, with typical temperatures of $\sim$ 100 K, volume densities $\sim$ 20 - 25 cm$^{-3}$, and volume filling factor  f $\sim$ 2 to 4\%. This phase is most readily traced by HI measured in absorption.

\item {\it  The  warm neutral medium (WNM)}, which provides the bulk of the HI seen in emission, with typical temperatures  $\geq$ 6000 K, volume densities $\sim$ 0.3 cm$^{-3}$ and volume filling factor  f $\geq$ 30\%.

\item {\it  The warm ionized medium (WIM)} with  T $\sim$  8000 K, n $\sim$  0.03 - 0.3 cm$^{-3}$, and f $\geq$ 15\%,  ionized gas associated with HII regions, but also diffuse filling a considerable fraction of the ISM.

\item {\it  The hot ionized medium (HIM)} with T$\sim$ 10$^6$ K, n$\sim 10^{-3}$ cm$^{-3}$ and  f $\leq$ 50\%. This phase of hot gas is produced by supernova explosions,  has a long cooling time and consequently a large fraction of the ISM is filled with this ``coronal" gas. It can be viewed as the accumulated superposition of dissipated SNRs interiors integrated over the lifetime of the Galaxy \citep{Mckee1977}.
\end{itemize}

 The cold, warm and hot phases are in global pressure equilibrium, while  the molecular material is mostly confined to  clouds which are held together by gravitation. The filling factor for each of the phases is highly uncertain, as is the topology of the ISM. In general, the ISM has structures on all scale lengths, from smaller than 1 pc to larger than 1000 pc. Through HI observations of the Galaxy it has been known that the ISM is pervaded by holes surrounded by  shells and supershells \citep{Heiles1984, McClure2002, Suad2014}, as the result of individual or collective action of  stellar winds of massive stars and  supernova explosions. The interior of these bubbles  is  filled with HIM. 

Molecular gas, on its side, is mostly ($\sim$ 90\%) assembled in massive structures  distributed in large clumps (GMCs, the giant molecular clumps with masses of $\sim 10^4 - 10^6$ M$_\odot$, diameters $\sim$ 50 pc and average densities $n_{H_2}\sim 100 - 300$ cm$^{-3}$),  clouds (M $\sim 10^4$ M$_\odot$, diam $\sim$ 5 pc,  $n_{H_2}\sim 300$ cm$^{-3}$), and condensed cores (possible birth sites of new stars, with M $\sim 10^3$ M$_\odot$, diam $\sim$ 2 pc,  $n_{H_2}\sim 10^3$ cm$^{-3}$) \citep[e.g.][]{Williams2000}.  They are principally compose of  by molecular hydrogen and dust and, in very little quantities,  by some of the $\sim$ 140 molecular species that have been identified in interstellar or circumstellar gas.

\subsection {Consequences of the interaction:}

 As expected, the expanding SNR shock front undergoes a different evolution according to the phase of the ISM that it encounters in its expansion, and the   dynamical evolution of a SNR is modified with respect to the simple description presented in Section \ref{sec:theory}. Given a constant pressure in the SNR and that the postshock pressure scales with $\rho_0 v_s^2$ (where $\rho_0$ is the ambient density and $v_s$ the shock velocity) it is expected  that the SN rapidly expands in the lowest density phase, while dense clouds are slowly crushed by lower velocity shocks. The  energy can be  conducted from the hot gas filling the SNR to embedded clouds, leading to their evaporation.  A complete treatment of the expressions that govern the expansion of a SNR in these cases can be found in \citet{Tielens2005}. In the end, the exchange of mass, energy, and momentum between the different phases govern the dynamical evolution of the SNR, and the appearance of the SNR in the different spectral regimes  reflects the structure of the surrounding medium.   

 Due to the short lifetimes of massive stars, most core-collapse SNe (resulting from explosions of types SNII, SNIb and SNIc)  are located close to the molecular concentrations where the precursors were born. Therefore a large percentage (as high as $\sim$ 75\%) of the Galactic SNRs are expected to be interacting with MCs. One important consequence of these interactions is that because of the compression and high temperatures of the SNR shocks propagating inside MCs, some chemical reactions otherwise not possible, can occur, creating new molecular species \citep[e.g.][]{Tielens2005}. For example, a recent work by \citet{Dumas2014} reported the detection of  SiO emission triggered by the passage of the W51C SNR shock. 

\citet{Slane2015} describe the X-ray and \gamray\ signatures of the interaction of SNRs with molecular clouds  and summarize the different criteria that can be used to establish  the existence of a  physical relation between an SNR and a cloud seen in projection against the remnant.  To unambiguously establish whether an SNR is physically associated with an interstellar cloud, removing confusion introduced by unrelated gas along the line of sight, is not trivial and usually requires of several distinct criteria to demonstrate physical interaction. Basically, morphological traces along the periphery of the SNRs, such as arcs of gas surrounding parts of the SNR or indentations in the SNR outer border encircling dense gas concentrations. Usually, such features are indicating that a dense external cloud is  disturbing an otherwise spherically symmetric  shock expansion. These initial signatures need to be confirmed with more convincing, though more rare, features like  broadenings, wings or asymmetries in the molecular lines spectra \citep{Frail1998}, high  ratios between molecular  lines of different excitation state  \citep{Seta98}, detection of near infrared H$_2$ or [Fe II] lines \citep[e.g.][] {Reach2005}, peculiar infrared colors \citep[e.g.][]{Castelletti2011a}, and the presence of OH (1720 MHz) masers.  These masers were originally detected by \citet{Goss1968} in some SNRs, but their importance as proof of SNR/MC interaction was only recognized almost 30 years later \citep{Claussen97, Frail1998}, and their detection became the most powerful tool to diagnose SNR/MC interaction. In addition to be an ideal tracer of interaction, observations of OH (1720 MHz) masers provide an accurate estimate of the magnetic field intensity of the post-shock gas in SNRs via Zeeman splitting \citep[e.g.][]{Brogan2000}. 

Based on a combination of different techniques, Jiang catalogued   a list of  $\sim 70$ Galactic SNRs candidate to be physically interacting with neighbouring MCs\footnote{http://astronomy.nju.edu.cn/$\sim$ygchen/others/bjiang/interSNR6.htm}, of which 34 cases are confirmed on the basis of  simultaneous fulfillment of various criteria,  11 are probable, and 25  are classified as possible and deserve more studies \citep{Chen2013}.

The basic method of investigating cases of SNR/ISM interaction is to survey  the interstellar medium in a field around the SNR using different spectral lines, from the cold, atomic hydrogen emitting at $\lambda$21 cm to the dense shielded regions of molecular hydrogen emitting in the millimetric and infrared ranges. The molecular gas is frequently  studied through  CO  observations (usually the transition $^{12}$C$^{16}$O~J: 1-0 at $\lambda$ 2.6 mm), since this molecule  radiates much more efficiently than the abundant H$_2$ and can be detected easily. Later, the observed CO intensity has to be converted into total H$_2$ gas mass.  According to the standard methodology, a simple relationship can be established between the observed CO intensity and the column density of molecular gas, such that N(H$_2$) = XCO $\times$ W(CO ~J: 1-0), with the column density, N(H$_2$) in cm$^{-2}$ and the integrated line intensity W(CO)  in  K km s$^{-1}$.  \citet{Bolatto2013} present a complete review of the theory,  techniques, and results of efforts to estimate XCO, the ``conversion factor"  from CO into H$_2$    in different environments. In the Milky Way disk, the representative XCO values varies between 0.7 and 2.8 $ \times 10^{20}$ cm$^{-2}$ (K km s$^{-1})^{-1}$ depending on the technique applied for the estimate. A value XCO = $2 \times 10^{20}$ cm$^{-2}$ (K km s$^{-1})^{-1}$  with 30\% uncertainty is recommended by \citet{Bolatto2013}.  

In HI (absorption and emission methods), numerous studies around SNRs have been carried out using single-dish and interferometric radiotelescopes,  looking for traces of disturbances in the interstellar gas distribution caused by the SNR or its stellar progenitor  \citep[e.g.][]{Dubner1998, Reynoso1999, giacani2000, Velazquez2002, Koo2004, Paron2006, Leahy2008, Lee2008, Park2013}.
An interesting approach using atomic gas studies was  recently used in the historic remnant of SN 1006 by \citet{Miceli2014}, where an important connection between shock-cloud interaction and particle acceleration  was demonstrated based on the comparison of X-ray with HI data.  

Many dedicated studies were conducted towards several Galactic SNRs in different  molecular transitions \citep[e.g.][etc.]{ReachRho99, Moriguchi2001, Dubner2004, Moriguchi2005, Reach2005, Jiang10, Paron2012, Li2012, Hayakawa2012, Fukui2012, Gelfand13, Kilpatrick2014}. Of particular importance have been the molecular gas investigations around the SNRs IC443, W28 and W44, which have been the subject of many observational studies.  These were the first SNRs where the OH(1720 MHz) masers were observed. In the recent years, these sources became especially notorious as they are good examples of Galactic \gamray\ sources  detected in the TeV range as seen by H.E.S.S. and other Cerenkov telescopes, and in the GeV range by the Fermi and AGILE satellites.

\begin{figure}
\centering
\includegraphics[width=1.0\textwidth]{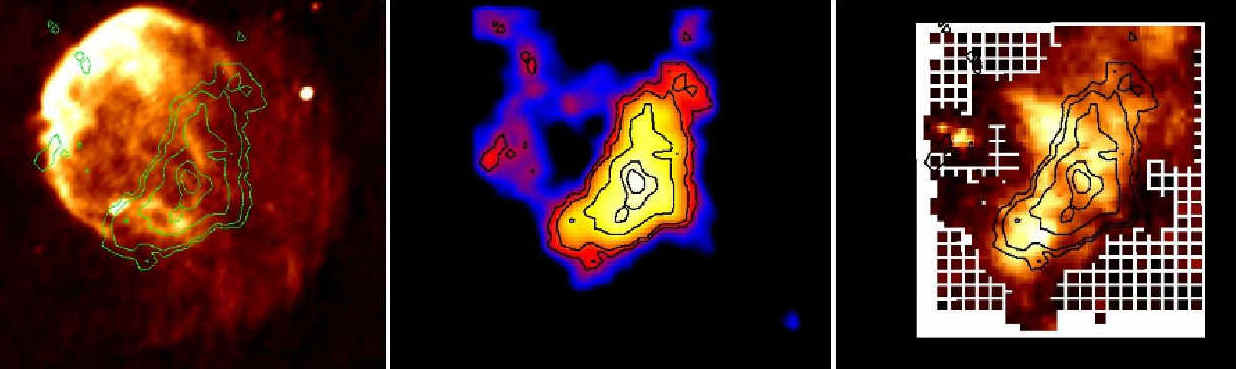}
\caption{The SNR IC443: {\it left:} radio-continuum at 330 MHz  \citep[in greys, from][]{Castelletti2011a} with \gamray\ emission from VERITAS (contours), {\it middle: } VERITAS image of very-high-energy (VHE, E $\geq$ 100 GeV) $\gamma$-ray emission \citep{Acciari09}; {\it right:} Molecular CO emission \citep [in greys, from][with \gamray\ emission overlapped in contours]{Zhang2010}}
\label{IC443-comparison}       
\end{figure}

The SNR IC443 (G189.1+3.0), because of its location in a relatively confusion free region of the outer Galaxy, is a text book case to analyze shock chemistry, and as such it has been thoroughly studied in many molecular transitions. From the first identification by \citet{Cornett1977},  tens of works investigated the chemical and physical transformations introduced by the strong SNR shocks on the surrounding molecular clouds \citep[see][for an updated summary of molecular studies towards IC443]{Kilpatrick2014}. The morphology consisting of two semi-circular shells with different radii is an indication of expansion in an environment with a marked density contrast.  IC443 is also an excellent example where the interaction of the SNR with a molecular cloud probably gave origin to the \gamray\ emission  through a hadronic mechanism. In effect, an excellent concordance was demonstrated by \citet{Castelletti2011a} by confronting  the  VERITAS very high energy \gamray\ radiation \citep{Acciari09} with the $^{12}$CO J=1-0 cloud \citep{Zhang2010} (Fig. \ref{IC443-comparison}), and the region where the radio-continuum emission of IC443 shows flat spectral indices not caused by thermal absorption (see Section \ref{sec:spectra}). 

\begin{figure}[ht!]
\includegraphics[width=0.4\textwidth]{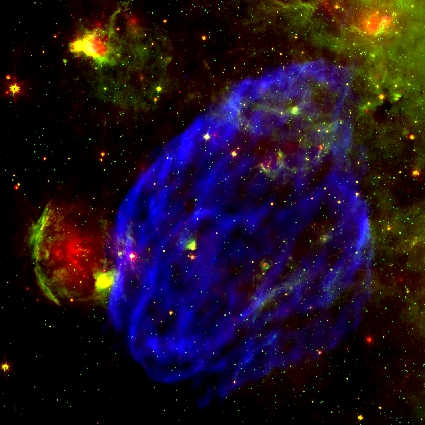}\hfill~
\includegraphics[width=0.6\textwidth]{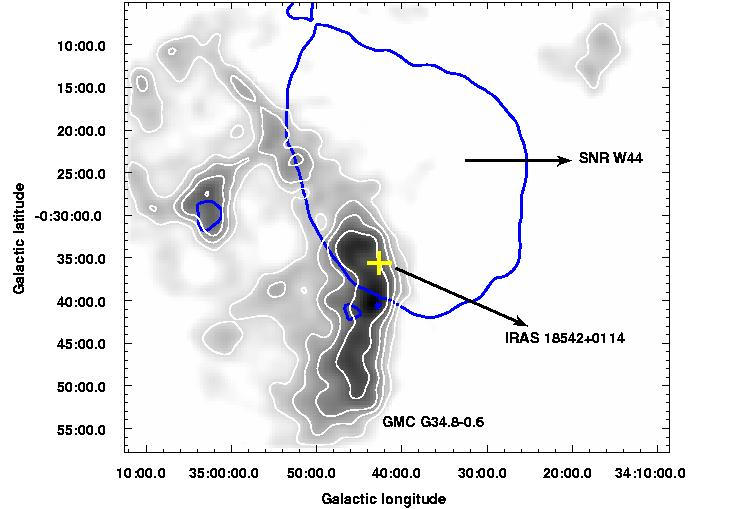}
      \caption{{\it Left:} Radio continuum emission of the SNR W44 at 324 MHz (VLA image, in blue) with the  infrared emission in the SNR region from {\it Spitzer} data at 8 $\mu$m (in green) and at 24 $\mu$m (in red) \citep[from][]{castelletti2007}. The yellow spot in the southeast corresponds to the HII region G034.8-0.7, an area where there is active star formation. {\it Right:} $^{13}$CO J = 1-0 distribution  showing the giant molecular cloud GMC G34.8-0.6 interacting with the SNR W44. The cross shows the position of the infrared source IRAS 18542+0114. The blue contours trace the silhouette of SNR W44 displayed in Galactic coordinates \citep[from][]{Paron2009}}
\label{W44}
\end{figure}

The SNR W44 (G34.7-0.4)  (Fig. \ref{W44}) is another Galactic remnant whose interaction with a cloudy ambient medium has been deeply investigated, at first motivated by its morphology of a rather distorted shell  with a significant flattening  along the eastern edge, suggestive of an encounter with dense ambient medium. Since the first HI study by \citet{Sato1974}, reporting the presence of a dense cold cloud in coincidence with W44, many observations have been performed in different molecular and IR lines \citep[e.g.][and references therein]{Dickel1976, Wootten1977, Seta2004, Paron2009, Sashida2013, Anderl2014}. The IR data showed that this SNR expands in a dense medium with n $\sim$ 100 cm$^{-3}$ \citep{Reach2005}. \citet{Giuliani2011} used constraints set by  CO data from the NANTEN Observatory, the radio spectrum as obtained by \citet{castelletti2007} and  optical data reported by \citet{Giacani1997}, to demonstrate that the   \gamray\ emission detected with AGILE in the energy range 400 MeV $-$ 3 GeV, was consistent with hadron-dominated models. The possibility of star formation  in the region of strong interaction between the expanding SNR shock and the adjacent molecular cloud (where there are IR sources with spectral characteristics of young stellar objects), is discussed below in Section 7.2.

\begin{figure}[ht!]
\includegraphics[width=0.35\textwidth]{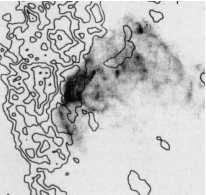}\hfill~
\includegraphics[width=0.65\textwidth]{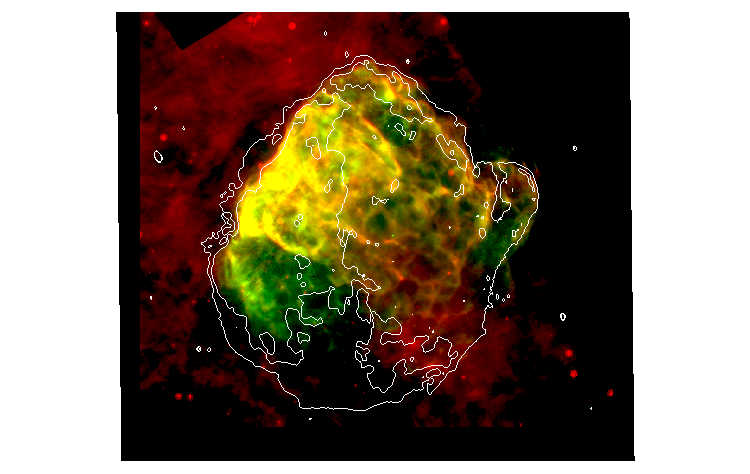}
      \caption{{\it Left:} Distribution of the HI emission (in contours) around the SNR Puppis A, overlapping the ROSAT X-ray emission (in greys) as taken from \citet{Petre1982}. Image extracted from \citet{Reynoso1995}. {\it Right:} Two color image of Puppis A comparing the {\it Chandra} and {\it XMM-Newton} X-ray emission in the 0.7-1.0 keV band (in green) \citep{Dubner2013}, with IR emission as observed with Spitzer at 24$\mu$m (in red) \citep{Arendt2010}. The white contours show radio-continuum emission at 1.4 GHz  \citep[from][]{Castelletti2006}}
\label{puppis-x-radio}
\end{figure}

 Another interesting observational example of interaction can be found in the SNR Puppis A (G260.4-3.4),  where CO and HI observations carried out by \citet{Dubner1988} first revealed  that the SNR is  expanding in an inhomogeneous environment, with clouds  towards  the east and northeast borders. Later, \citet{Reynoso1995} explored this SNR in HI with high angular resolution using the VLA (Fig. \ref{puppis-x-radio} {\it Left}) revealing the distribution of a pre-existing interstellar HI cloud closely following the borders of the SNR shell as seen in X-rays.  Inside the SNR, \citet{Hwang2005} investigated through X-ray images and spectra  the most prominent spots of cloud-shock interaction (mainly the bright eastern knot, known as the BEK), revealing what has been the  first 
X-ray identified example of a cloud-shock interaction in an advanced phase. From new CO observations, \citet{Paron2008} showed evidence of what has been left of the molecular clump engulfed by the SN shock and  that currently evaporates emitting in X-rays. Another interesting conclusion is that to the eastern boundary where the atomic gas is in touch with the SNR, the molecular emission is detached,  indicating that the precursor radiation has dissociated the  molecules of the adjacent cloud. It is also noticeable from Figure \ref {puppis-x-radio} {\it Right} that in X-rays Puppis A has a  ``cellular"  filamentary structure with a honeycomb appearance, confirming that it expands in a rich and complex environment. In addition, Figure \ref {puppis-x-radio} {\it Right}, showing the location of the radio border running outside of the sharp X-ray and IR limbs, is clear evidence that the radio radiation is the best indicator of the true position of the shock front.

\subsection{ Can SNRs trigger the formation of new stars?}

 Since \citet{Opik1953}'s early idea that ``the death of a star in a
supernova explosion may lead to the birth of a great number of new stars",
predicting that a ring of newborn stars would condense out of the
expanding shell, it became a paradigm that one of the consequences of
SNR/ISM interactions can be the formation of new stars. \citet{Vanhala1998}
carried out numerical studies, concluding that shock waves with velocities
in the range of 25 $-$ 40 km s$^{-1}$ are capable of triggering collapse in
molecular cores immersed in clouds interacting with SNRs. 
\citet{Melioli2006} set constraints on the SNR radius and molecular gas
density for which star formation is allowed. If the SNR-cloud interaction
is too strong, the cloud is completely destroyed; if it is too weak, the
molecular core never collapses. In summary, the theoretical conclusion is that if favourable conditions are
given, triggered star formation can occur as a consequence of SNR/ISM
interaction.

 \citet{Wootten1978} conducted one of the very first searches of star formation triggered by a SNR. From the analysis of an infrared object in
a cloud near the SNR W44, the author concluded that it was an evidence of star
formation estimulated by the expansion of the SNR. Since then, several searches have been
carried out, including one based on multi-wavelegth data towards W44, where  \citet{Paron2009} demonstrated that the star formation was related to the nearby HII region G034.8-0.7 and not to the SNR. Other
 SNRs interacting with molecular clouds with indicators of active star formation have also been investigated, including W30
\citep{Ojeda-May2002}, G54.1+0.3 \citep{Koo2008}, G357.7+0.3
\citep{Phillips2009}, IC443 \citep{Odenwald1985, Xu2011}, G18.8+0.3 \citep{Dubner2004, Paron2012}, etc. In all
cases, even when the location of protostellar objects and young stellar
object (YSO) candidates immersed in shocked molecular clouds were
very promising, after comparing the characteristic timescale of star
formation with the age of the SNRs, the general conclusion is that the
stellar formation started before the SN explosion.

\citet{Desai2010} carried out the most complete and homogeneous search for
star formation related to SNRs by examining the presence of YSOs and molecular clouds in the environs of 45 SNRs in the Large
Magellanic Cloud. After a very detailed analysis, based on different arguments
(positions, densities, timescales, etc.) the conclusion is that there is
none evidence of SNR-triggered star formation in the LMC.

In conclusion, star formation is frequently seen near supernova remnants,
but such physical association does not necessarily imply a causal
relationship. Massive stars that end their lives exploding as SNe are
formed in clusters or OB associations where the formation of new stars may continue and
propagate outward for a prolonged period of time. It is then not surprising
that core-collapse SNe are near young stars in star-forming environments.
But apparently the SNR shocks result in an increment of turbulence that is
not compatible with star formation. In fact, the agitation may be so
violent that disperses the material, hindering further star-forming
activity. This is an important open field of research that demands further theoretical and observational studies.

\section{ Comparison of radio emission with emission in other spectral ranges}
\label {sec:comparison}

The emission in different spectral regimes traces material with different physical conditions. Briefly, optical filaments observed in most SNRs arise from shocked interstellar medium that is cooling radiatively, while in a few remnants the optical emision  includes oxygen rich filaments,  fragments of nearly pure ejecta launched from the core of the progenitor star during its explosion, or can be dominated by Balmer lines when a fast shock (velocities higher than $\sim$ 200 km s$^{-1}$) enters into partly neutral interstellar gas.  The thermal X-ray emission contains  essential information regarding the temperature, composition, distribution and ionization state of the material synthesized and ejected in SN explosions, and of the ambient matter swept-up by the supernova shock. If the origin of the X-ray emission is synchrotron, then its study can set powerful constraints on the role of the SN shocks in the production of cosmic rays. Infrared emission mainly marks the location of shock-heated dust. In summary, the body of multispectral observations is not only useful to understanding properties of the SNRs and their precursors, but also to explore the interstellar gas.

The search for correlation between radio and IR emissions is a powerful tool for distinguishing thermal from non-thermal emission. The combination of the VLA survey of the Galactic Plane at 330 MHz with the Midcourse Space Experiment (MSX) at 8$\mu$m, served to discover 31 new SNRs and 4 candidates in the inner Galactic plane region, where the diffuse synchrotron emission and thermal HII regions cause more confusion. The SNRs are anti-correlated with dust emission, while HII regions are invariably surrounded by a shell of bright 8$\mu$m emission \citep{Brogan2006}. Also, this methodology has been applied to provide accurate estimate of the integrated radio flux density in remnants located in complex regions of our Galaxy where the contamination with thermal structures is high, as for example in the SNRs RX J1713.7-3946 \citep{Acero2009} and G338.3-0.0  \citep{Castelletti2011b}. Besides, the comparison with radio emission from SNRs has been one of the prime methods to identify the nature of several \gamray\ sources detected in the GeV and TeV ranges (more than twenty Galactic TeV sources are related with SNRs).

 Nowadays, the study of almost all SNRs is tackled from a multi-wavelength approach. In what follow, we describe some sources in which the comparison of radio emission with emission in other spectral windows unveiled interesting results.

\begin{figure}[ht!]
\centering
\includegraphics[width=0.8\textwidth]{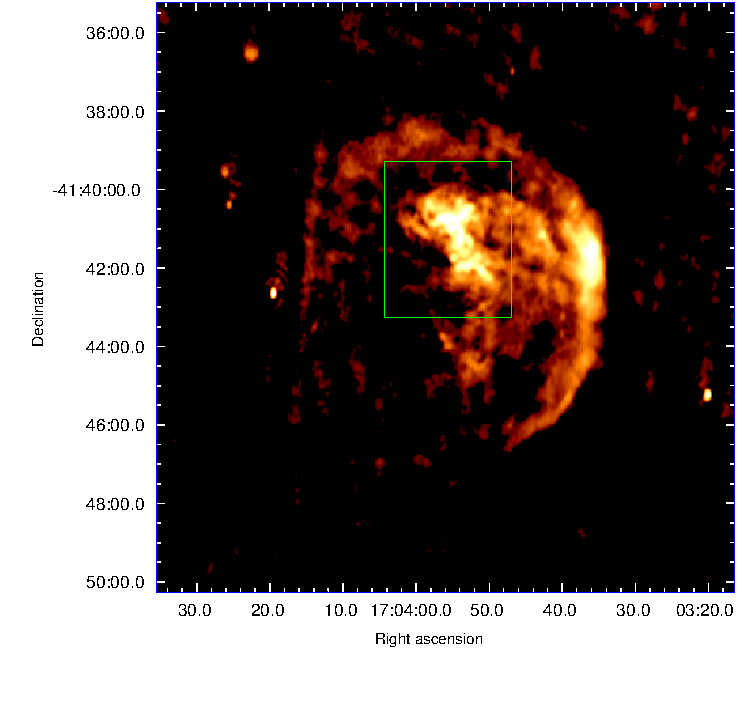}~\hfill
      \caption{Radio image of G344.7-0.1 at 1.4 GHz taken from \citet{giacani2011}. The green square shows the location of the radio nebula near the geometric center of the remnant}
\label{g344-1.9}
\end{figure}

{\it G344.7-0.1} This SNR is a clear example where the comparison between  radio, IR at 24$\mu$m and X-ray images  elucidated the nature of a radio nebula located near the center of the remnant (Fig. \ref{g344-1.9}). From a radio spectral study carried out in this region of the remnant using the Very Large Array (VLA, NRAO) and the Australia Telescope Compact Array (ATCA) data at 1.4 GHz and 5 GHz, \citet{giacani2011} determined a mean radio spectral index of $\alpha \sim$ 0.3 for the nebula, a value compatible with those of radio PWNe. The combination of the emission in the three mentioned spectral ranges allowed the authors to rule out a PWN origin, concluding that the central bright radio feature is probably the result of strong shocks interacting with dense material. This interaction enhanced the  infrared emission from shocked dust  and favored particle acceleration, resulting in a flatter radio spectrum. 

\begin{figure}[ht!]
\centering
\includegraphics[width=0.5\textwidth]{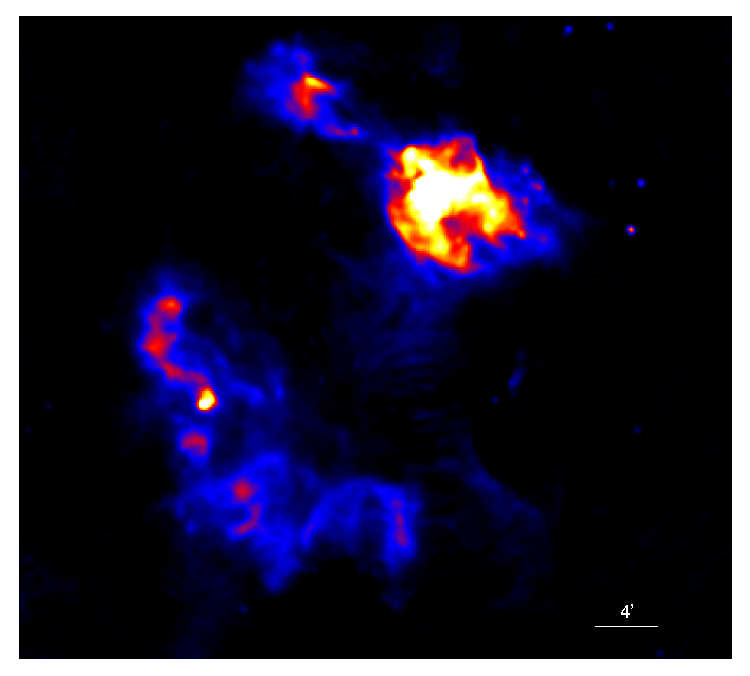}
\includegraphics[width=0.27\textwidth]{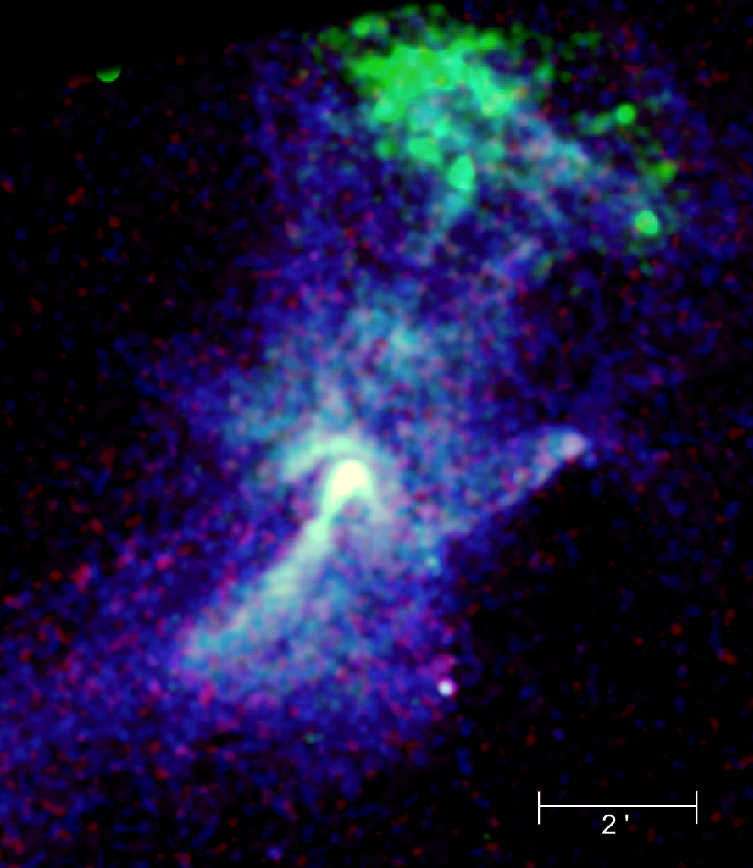}
      \caption{{\it Left:} Radio continuum image of the SNR G320.4-1.2 at 1.4 GHz taken from \citet{dubner02}. {\it Right:} X-ray image by {\it Chandra} \citep{Gaensler2002}. Credit Image NASA/MIT/B. Gaensler et al. }
\label{g320}
\end{figure}

{\it G320.4-1.2 (MSH 15-52)}  A complex interacting system that has been observed throughout the electromagnetic spectrum is formed by the SNR G320.4-1.2, the energetic pulsar PSR B1509-58 and its PWN, and the HII region RCW 89 (Fig. \ref{g320}).  SNR G320.4-1.2 has an unusual radio appearance consisting of two distinct components:  towards the northwest, a  bright centrally concentrated source with a ring of radio clumps in coincidence with the optical nebula RCW 89,  and towards the southeast a fainter partial shell (Fig. \ref{g320} {\it Left)}. The pulsar PSR B1509-58, detected in radio, X-rays and in \gamray\, is located near the center of G320.4-01.2. In the X-ray domain, this system has a complex picture with a different morphology (Fig. \ref{g320} {\it Right}). The emission is dominated by a jet-like structure emerging from  the pulsar,  the PWN, and  extended emission, thermal in origin  coincident with the northern radio component of the SNR and with the HII region RCW 89. \citet{gaensler99} proposed that this complex system can be explained as the result of a low-mass or high-energy explosion occurring near one edge of an elongated low density cavity, which was confirmed by observations of the neutral gas in the region by \citet{dubner02}. The pulsar appears to be generating twin collimated outflows, the northern part of which interacts with the SNR producing the collection of the clumps observed in radio and in the X-ray band. \citet{Gaensler2002}  later reported the discovery of the long-sought  PWN on the basis of comparison of new {\it Chandra} X-ray data with radio morphology.

{\it G292.0+1.8} This remnant is another interesting source that has been subject  to multi-wavelength observational campaigns. It is a  textbook example because it has all the properties of a core-collapse SNR: a central pulsar (PSR J1124-5916)  feeding a PWN, metal-rich ejecta,  shocked CSM, and a blast wave. For a summary of previous observations, see \citet{ghavamian12}. Briefly, it is characterized by a remarkable network of filaments seen in  optical and in X-rays, which are fragments of supernova ejecta  \citep{winkler06, park02, park04, park07}. The X-ray image also reveals bright  filaments running from east to west, named the  ``equatorial belt", probably shocks propagating in circumstellar material \citep{park02}. In the radio band, a mutifrequency study of the remnant performed with ATCA  at 20, 13 and 6 cm  by \citet{gaensler03}, revealed  the presence of a polarized and flat-spectrum bright central core, representing the radio PWN, surrounded by a circular fainter plateau with steep-spectrum which represents the SNR shell. Curiously, a series of radial filaments with much flatter spectrum are located over the plateau. The comparison with the X-ray emission allowed  \citet{gaensler03} to propose that Rayleigh-Taylor instabilities near the SNR contact discontinuity originated the formation of such filaments  \citep[Fig.3 in][]{gaensler03}.

\begin{figure}[ht!]
\centering
\includegraphics[width=0.6\textwidth]{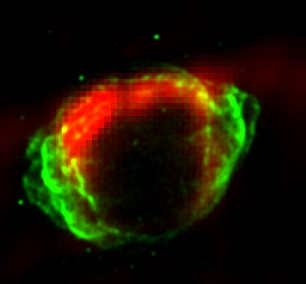}
\caption{ {\it Chandra} image of G1.9+0.3 (in green) \citep{Borkowski2013} superimposed with VLA radio image at 4.9 GHz (in red) as reprocessed from archival VLA data by the authors of this paper}
\label{g1.9}
\end{figure}

{\it G1.9+0.3} It is the youngest remnant detected in the Milky Way  (age $\sim $150 yr), opening as such the opportunity to study a Galactic SNR on its very early development. The  morphology of G1.9+0.3 can be described as an almost complete shell in radio and in X-rays \citep{Green2008, Reynolds2008}, but the comparison between the distribution of the emission in both spectral ranges  revealed a peculiar anti-correlation. While the radio remnant is clearly brighter  along the northern border, the X-rays has a notable bilateral east-west symmetry including two extensions (``ears") not detected in the radio band (Fig. \ref{g1.9}). It is one of the few shell SNRs with an X-ray spectrum dominated by synchrotron emission \citep{reynolds09}. But this is not the only origin of the X-ray emission, \citet{borkowski2010} reported the presence of lines of Fe, Si, and S in small regions of the northern limb with spectroscopic velocities of about 14000 km s$^{-1}$ and a line at 4.1 keV, identified as due to $^{44}$Sc (the first firm detection of this element in a SNR).  Based on the presence of Fe lines, high velocities, absence of PWN, and bilaterally symmetric non-thermal X-ray emission as in SN 1006, \citet{Borkowski2013} suggest a Type Ia origin for G1.9+0.3. 

\section{Supernova Remnants in the Magellanic Clouds}
\label {sec:magellanic}

The Magellanic Clouds  are an ideal laboratory to investigate the properties, both global and individual, of SNRs across the whole electromagnetic spectrum. The fact that the distances are known (49 kpc for the LMC and 59 kpc for the SMC), that the LMC is nearly face-on, so that all remnants are nearly co-distant, and that even for the rounder SMC there is a range of only $\pm$ 5 kpc with respect to the mean distance, allows for accurate estimation of physical parameters of the remnants. At the same time, the proximity permits radio observations with  good angular resolution revealing the structure of individual objects and facilitating  accurate multi-spectral comparisons.

One of the very first identifications of a  SNR in the Magellanic Clouds was reported by \citet{Mathewson1963} based on radio observations carried out with the 210-ft. steerable reflector at the Australian National Radio Astronomy Observatory. It was followed by several searches using the Molonglo Observatory Synthesis Telescope (MOST) that permitted the identification of  new members based on the combination of radio observations with optical and X-ray data. \citet{Mathewson1983} compiled a catalogue of SNRs in the  Magellanic Clouds with 25 confirmed remnants in the LMC and 6 in the SMC. Later on, the search and studies of SNRs in our neighbouring dwarf galaxies  took advantage of the Australia Compact Telescope Array (ATCA)  that improved the angular resolution by factors between 5 and 35 compared with the best previously available radio data. Several works reported discoveries of new members and detailed studies of already known SNRs \citep[e.g.][etc.]{Amy1993, Manchester1993a, Manchester1993b, Dickel1993,Dickel1995, Dickel1998}

\citet{Badenes2010} compiled from the existing literature a catalogue of multi-wavelength observations with 77 confirmed SNRs in the Magellanic Clouds (54 in the LMC and 23 in the SMC), arguing that this list comprises a fairly complete record of SNe that exploded over the last $\sim$ 20 kyr. After Badenes published the catalogue,  at least seven new  cases were confirmed in the LMC \citep{grondin2012, maggi2012, kavanagh2013, bozzeto2013, maggi2014}. Of all the 84 SNRs identified to date, only  4 of them (SNR J005.9-7310  in the SMC, and  MC SNR J058-6830, MC  SNR J0511-6759 and MC SNR J0517-6759 in the LMC) have not been detected in the radio band, possibly because they are old and faint \citep{maggi2014}. 

\citet{Badenes2010}  carried out statistical studies of the SNRs in the Magellanic Clouds concluding that the size distribution of remnants is approximately flat with a cutoff at r$\sim$ 30 pc. The authors propose that  most of the SNRs are in  the  Sedov phase of evolution, quickly fading below detection as soon as they reach the radiative stage.  

Among all  SNRs in the  Magellanic CLoud, the radio remnant of SN 1987A deserves special attention because it offered the possibility of tracing in great detail its temporal evolution. In what follows, the main aspects of this source are summarized.

\paragraph {The radio remnant of the SN 1987A}

SN 1987A was the first  naked-eye SNe event since the invention of the telescope. It exploded in 1987 February 23 in the Large Magellanic Cloud. Briefly, the progenitor of SN 1987A, Sk-69$^{\circ}$202 with an initial mass estimated in $\sim$ 20 M$_{\odot}$, is believed to have evolved from a red supergiant (RSG) into a blue supergiant (BSG) approximately 2$\times 10^4$ yr prior to the explosion \citep{crottsHeathcote2000}. Optical imaging revealed a complex 
circumstellar medium (CSM) surrounding the explosion, consisting of a triple-ringed  structure \citep[and reference therein]{burrows95}. The two outer rings appear to be the cap of an hourglass-shaped structure enveloping the SN itself. The inner ring, also referred to as the equatorial ring, is believed to represent an equatorial density enhancement in the CSM, located at the interface between a dense wind emitted from an earlier red-giant phase of the progenitor star and a faster wind emitted by the star in more recent times.

In the radio band, the emission from the SN 1987A was detected at 843 MHz with the MOST two days after the SN event. This emission reached its maximum value of around 140 mJy 4 days after the explosion and  then decayed rapidly to become undetectable less than a year later \citep{turtle87}. This radio outburst has been explained as the consequence of the rapidly moving, low density BSG wind, which produced only a short-lived period of radio emission when hit by the SN shock \citep{storey87, chevalier98}. 

Approximately 1200 days after the explosion, radio synchrotron emission was again detected, in this case by both ATCA and MOST \citep[][respectively]{staveleySmith92, ball01}, marking the birth of the radio remnant.  From the first radio detection, the radio emission has been regularly monitored every 1-2 months at 1.4, 2.4, 4.8 and 8.6 GHz using various array configurations of ATCA  \citep[and reference therein]{manchester02, staveleySmith07, zanardo2010}. At all these frequencies, the flux density has been steadily increasing over time and from $\sim$ day 3000 after the explosion is growing exponentially, which is attributed to increasing efficient particle acceleration processes  \citep{zanardo2010}. On the other hand, the highest angular resolution observations ($\sim 0^{\prime\prime}.1$) of the SNR  obtained with the Australian Large Baseline Array (LBA) at 1.4 and 1.7 GHz revealed two extended lobes with an overall morphology in good agreement with that at lower angular resolutions. In addition,  small-scale structures were found in the brightest regions in both lobes \citep{tingay09, ng11}. 

\begin{figure}[ht!]
\includegraphics [width=0.9\textwidth] {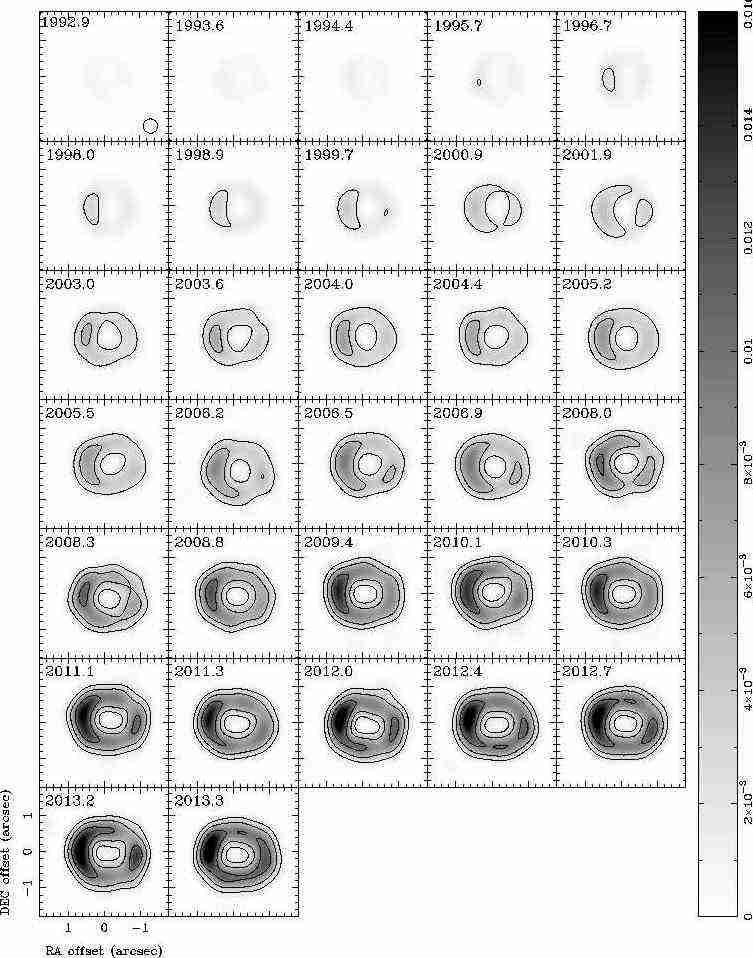}
      \caption{9 GHz images of SN 1987A in the interval 1992-2013 obtained with a FWHM of 0$^{\prime\prime}.4$ \citep[figure provided by][]{ng2013}}
\label{sn1987a}
\end{figure}

Recently, \citet{ng2013} reported on the study of the evolution of the radio morphology of the remnant of the SN 1987A  covering the period January 1992 - May 2013 (day 9568 after the explosion). The data were acquired with ATCA at 9 GHz with an angular resolution of 0$^{\prime\prime}.4$. The remnant presents a double lobe ring,  with an asymmetric surface brightness distribution being the eastern lobe brighter than its western counterpart (Figure~\ref{sn1987a}). From the analysis of this database the authors point out that from day 7000 the asymmetry began to decline, such that the overall geometry is evolving towards a ring structure, suggesting that the remnant has entered in a new evolutionary stage where the forward shock has fully engulfed the entire inner ring and is now interacting with the densest part of the circumstellar medium. 

Observations at higher frequencies (18, 36, 44 and 94 GHz) undertaken with ATCA \citep{manchester05, potter09, lakicevic12, zanardo2013}, followed by observations at 110, 215, and 345 GHz with the Atacama Millimeter/Submillimeter Array (ALMA), show the same asymmetric double lobe morphology \citep{indebetouw2014, zanardo2014}. 
Also ALMA observations \citep{indebetouw2014} revealed the presence of one of the largest masses of cold dust (greater than 0.2 M$_{\odot}$) ever measured in a SNR. The dust emission is concentrated at the center of the remnant, and has probably  formed in the inner ejecta. 

The evolution of the radio spectral index has been tracked over the years. At present the mean spectral index estimated between the frequencies 18 and 44 GHz is $\alpha $ = 0.8, with values in the range 1.1 to 0.3 across the source. The brightest region on the eastern lobe is  slightly steeper than the mean value, while flatter values, between 0.5 and 0.3, are found toward the central and the center-north regions of the SNR \citep{zanardo2013}. The spectral index distribution of the synchrotron component was mapped in the frequencies range 44 to 672 GHz with data obtained with ATCA and ALMA instruments \citep{zanardo2014}. From radio to far infrared, the authors estimate a mean spectral index for the synchrotron main component of 0.73, while a value $\sim 0$ was estimated around the central region and between 0.1 and 0.4 in the western half of the remnant. \citet{zanardo2014} conclude that central region might be  a PWN in the interior of the remnant, while the steepening of the spectral indices over the eastern lobe might be indication of a local spectral break.

\section{Conclusions and prospects of studies with the new generation radiotelescopes}

Almost seven decades after the first detection of the radio emission coming from SNRs, great advances in the knowledge of their properties have been achieved. Detailed studies of the radio emission associated with SNRs have proven to be an excellent tool to recognize morphological characteristics, delineate the location of shock fronts, identify sites of interaction with dense clouds of the ISM,  locate places of particle acceleration, and discover PWNe  even in cases where the radio pulsar is not detected and the nebula is the only trace of its existence. They are also  powerful probes  of the magnetic field, providing information on the orientation and sites of higher compression.  Furthermore, the possibility of resolving individual radio structures down to spatial scales of arcsec or even sub-arcsec, like in the expansion studies performed in Tycho's and SN1006 SNRs or the current studies of the Crab Nebula, is a powerful resource to investigate the dynamical evolution of SNRs. 

Over the past years, key observations performed with single-dish telescopes and radio interferometers like ATCA, DRAO, VLA, VLBA, GMRT, etc. provided remarkable insights on radio SNRs. However, there still persist controversial issues and unsolved questions requiring further research  from observational as well as from theoretical approaches. In brief, particular issues  deserving further observational studies are: the deficiencies of the spectral database, the generally poor quality of spatially resolved spectral indices over wide frequency ranges,  the scarcity of well determined distances, the incomplete knowledge of extended old SNRs, the missing (young and old) Galactic remnants, the consequences at all levels of the shock/cloud interactions, the alignement and strength of the magnetic field, and where and how  particles are accelerated. From the theoretical point of view, the comprehension of SNRs would benefit from progress in the research of mechanisms of particle acceleration capable of explaining the variety of observed spectral indices, as well as from detailed modelling of  the dynamical evolution of SNRs expanding in inhomogeneous environments.

Other motivation to increase the quality and accuracy of radio measurements is to match the ongoing revolution in high energy (X-ray and \gamray) studies. It is well known that recent surveys of the Milky Way with space and ground-based \gamray\ detectors revealed hundreds of high-energy and tens of very-high-energy \gamray\ emitters representing several Galactic source populations, including numerous SNRs and PWNe whose emission mechanisms remain poorly constrained. The fundamental nature of many of the high energy sources could greatly benefit from radio observationes that are only recently becoming technically feasible, e. g. with the JVLA or ALMA.

Another area with pending studies is the investigation of faint SNRs, where the lack of precise information from radio data may affect the comprehension of  high-energy phenomena and, at large, the question of the contribution of SNRs to the overall flux of Galactic cosmic rays. This is, for example,  the case of the SNRs RX J1713.7-3946 and RX J0852.0-4622, two very interesting shell-type X-ray SNRs  with  bright \gamray\  emission (at 1 TeV RX J0852.0-4622 is  as bright as the Crab Nebula), but with poorly constrained radio parameters. These sources are radio weak, confused on parts with thermal emission and with significant Galactic background variation, characteristics that all together are detrimental to  the accurate estimate of radio parameters.  For cases like these, the new low-frequency synthesis arrays  like, low-frequency array (LOFAR) in Europe, the Murchison widefield array (MWA) in Australia, and  the  Low-Band JVLA system  in  the USA, are important new tools. Their good sensitivity and angular resolution is key to imaging and differentiating the non-thermal radiation (bright at low radio frequencies) from optically thick and thin thermal emission from HII regions (which are intense at high-frequencies).  Low radio frequency studies are also indispensable to improving the precision of the continuum spectra of all SNRs. Accurate low-frequency data anchor the spectrum, avoiding the uncertainties of extrapolation from higher frequency data which render them insensitive to curvature,  an important piece of information  to  test the predictions of particle acceleration theories.
In addition, sensitive high-resolution meter-wavelength studies of selected Galactic regions, have  already  proven to be extremely useful to discover new SNRs. The detection of many new SNRs  would alleviate the incompleteness of the current Galactic SNRs census with important implications for the SNe rate and energy input into the ISM. A deep meter-wavelength census of Galactic SNRs  with the JVLA is currently under way, as are LOFAR studies of selected SNR complexes.

In the other extreme of radio frequencies, observations of  radio-continuum emission of bright SNRs  in the tens and hundreds of GHz regime using ATCA, ALMA, and JVLA, among other facilities,  have come to fill in the gap between radio and IR, offering significantly improved constraints on SNRs spectral energy distributions.

The  new aperture-synthesis telescopes, either in operation, in commissioning, or in construction phase (EVLA, SKA and its precursors ASKAP in Australia, MeerKAT in South Africa,  LWA in the USA, and FAST in China, plus the already mentioned MWA) will also bring important advances in understanding the properties of the magnetic field in SNRs. They will provide high angular resolution and will be able to produce wideband multichannel data amenable to rotation measure (RM) synthesis. Wide area polarization surveys will be very helpful to establish the presence of magnetic field, its direction, strength and spatial scale in a large sample of Galactic SNRs and in the Magellanic Clouds.  In addition,  very long baseline interferometers,  like the European Network eEVN, are the appropriate instruments  to resolve the structure and investigate the expansion of young SNRs located in nearby galaxies other than the Magellanic Clouds, following the pioneering studies conducted by \citet{Pedlar1999} in M 82 and \citet{Bartel1994} and \citet{Marcaide1995} in M 81, and the more recent from \citet{Bietenholz2010a} in NGC 4449, \citet{Bietenholz2010b} in NGC 891, etc.

Another chapter connected with the investigation of radio SNRs that deserves special attention is the study of the atomic and molecular emission from the surrounding ISM. Such studies have become more and more important to understanding the properties of the SNRs across the whole electromagnetic spectrum. Large-scale HI surveys, like SGPS (Southern Galactic Plane Survey) in the southern hemisphere and VGPS (VLA Galactic Plane Survey) in the north, have been very helpful to explore in emission and absorption the characteristics of the ISM towards  a large number of SNRs close to the Galactic plane. Also intermediate resolution CO surveys, like the GRS (Galactic Ring Survey of $^{13}$CO in the northern sky) and the $^{12}$CO NANTEN 4-m dish observations in the southern sky, have been important to localize possible sites of SNR/MC interaction. For sources at higher Galactic latitudes, the atomic and molecular studies require dedicated  observations. The necessity, however, of more detailed studies of the surrounding ISM for a large number of SNRs is evident. An ample, good-quality atomic database will allow us to refine distance estimates and, particularly,  HI data acquired with very high angular resolution and sensitivity can be the way to detect the HI shell predicted to form behind the expanding shock front when radiative cooling is sufficient to recombine a detectable amount of atomic material. On the other hand, molecular data at sub-arcmin angular resolution, as can be obtained with single-dish of intermediate size, like the 12-m antenna of the future Argentina-Brazil radiotelescope Long Latin American Millimetre Array (LLAMA), will be an adequate step forward following the NANTEN southern sky survey. The data will have the sensitivity and angular resolution suitable to cover large areas identifying compact molecular clumps that might have been overrun by SNRs. 

 In view of the present radio astronomy landscape, with new instruments already in use or coming into operation in the next future, plus upgrading of existing instruments with  the latest technological developments,  the panorama of future research of radio SNRs is very promising.

\begin{acknowledgements}

We are very grateful to our colleagues  Namir Kassim, David Green and Gabriela Castelletti for the critical reading of this manuscript and useful comments. We thank CONICET (Argentina) for support through the grant PIP 0736/11 and to ANPCyT (Argentina) through grant PICT 0571/11. We have used images provided by  C. Brogan, F. Giordano, L. Ksenofontov, C.-Y. Ng, S. Pineault, W. Reich, S. Reynolds, and E. Reynoso with permission of the authors, to whom we thank. 

\end{acknowledgements}

%\bibliographystyle{aps-nameyear}
\bibliographystyle{aa}
\bibliography{bib_snrs-agosto}   % name your BibTeX data base

\nocite{*}

\end{document}